\documentclass[amsmath,amssymb,twocolumn,notitlepage,nofootinbib,superscriptaddress,floatfix,eprint]{revtex4-2}

\usepackage[utf8]{inputenc}
\usepackage[english]{babel}
\usepackage{graphicx}
\usepackage{float}
\usepackage{color}
\usepackage[usenames,dvipsnames,table]{xcolor}
\usepackage[normalem]{ulem}
\usepackage{amsthm}
\usepackage{eucal}
\usepackage{times,bbm,amsmath,amssymb}
\usepackage{epsfig,color}
\usepackage{xcolor}
\usepackage{bm}

\usepackage{float,siunitx}
\usepackage[caption = false]{subfig}

\usepackage{sidecap}
\usepackage[scaled=.8]{couriers}
\usepackage{pstricks}
\usepackage{multirow}
\usepackage{placeins}
\usepackage{relsize}
\usepackage{pst-grad,bm}
\usepackage{epigraph}
\usepackage{gensymb}
\usepackage{longtable}
\usepackage{soul}
\usepackage{ulem}
\usepackage{xr-hyper}
\usepackage{hyperref}
\usepackage{cleveref}
\hypersetup{
    colorlinks = true,
    allcolors=blue
}
\usepackage{layouts}
\usepackage{acronym}
\usepackage{physics}
\usepackage{easyReview}

\usepackage{amsmath,amssymb,amsfonts}
\usepackage{textcomp}
\usepackage{physics}

\usepackage[ruled,linesnumbered]{algorithm2e}
\usepackage{verbatim}
\usepackage{tabularx}

\usepackage{booktabs}

\newenvironment{protocol}[1][htb]
  { 
   \begin{algorithm}[#1]%
  }{\end{algorithm}}

\usepackage{microtype}
\microtypecontext{spacing=nonfrench}
\microtypesetup{
protrusion={true,nocompatibility},
activate={true,nocompatibility},
tracking=true,
kerning=true,
spacing={true}
}
\usepackage[all=normal,floats=tight,mathspacing=tight,wordspacing=tight,paragraphs=normal,tracking=tight,charwidths=tight,mathdisplays=tight,sections=normal,margins=normal]{savetrees}

\usepackage[style=american,autopunct=true]{csquotes}

\usepackage{placeins} 

\definecolor{blue}{rgb}{0,0,1}
\definecolor{grey}{rgb}{0.6,0.6,0.6}

\definecolor{myurlcolor}{rgb}{0,0,0.7}
\definecolor{myrefcolor}{rgb}{0.8,0,0}
\definecolor{purple}{RGB}{128,0,128}
\definecolor{ultramarine}{RGB}{63, 0, 255}
\definecolor{medblue}{RGB}{0, 0, 100}
\definecolor{googleblue}{RGB}{34, 0, 204}
\definecolor{panblue}{RGB}{0,24,150}
\definecolor{carmine}{RGB}{150, 0, 24}
\definecolor{gray}{RGB}{150, 150, 150}

\begin{document}

\title{Multi-client distributed blind quantum computation with the Qline architecture}

\author{Beatrice Polacchi}
\affiliation{Dipartimento di Fisica - Sapienza Universit\`{a} di Roma, P.le Aldo Moro 5, I-00185 Roma, Italy}

\author{Dominik Leichtle}
\affiliation{Laboratoire d’Informatique de Paris 6, CNRS, Sorbonne Université, 75005 Paris, France}

\author{Leonardo Limongi}
\affiliation{Dipartimento di Fisica - Sapienza Universit\`{a} di Roma, P.le Aldo Moro 5, I-00185 Roma, Italy}

\author{Gonzalo Carvacho}
\affiliation{Dipartimento di Fisica - Sapienza Universit\`{a} di Roma, P.le Aldo Moro 5, I-00185 Roma, Italy}

\author{Giorgio Milani}
\affiliation{Dipartimento di Fisica - Sapienza Universit\`{a} di Roma, P.le Aldo Moro 5, I-00185 Roma, Italy}

\author{Nicol\`o Spagnolo}
\affiliation{Dipartimento di Fisica - Sapienza Universit\`{a} di Roma, P.le Aldo Moro 5, I-00185 Roma, Italy}

\author{Marc Kaplan}
\email{kaplan@veriqloud.fr}
\affiliation{VeriQloud, 13 rue Victor Hugo, 92 120 Montrouge, France}

\author{Fabio Sciarrino}
\email{fabio.sciarrino@uniroma1.it}
\affiliation{Dipartimento di Fisica - Sapienza Universit\`{a} di Roma, P.le Aldo Moro 5, I-00185 Roma, Italy}

\author{Elham Kashefi}
\email{elham.kashefi@lip6.fr}
\affiliation{Laboratoire d’Informatique de Paris 6, CNRS, Sorbonne Université, 75005 Paris, France}
\affiliation{School of Informatics, University of Edinburgh, 10 Crichton Street, EH8 9AB Edinburgh, United Kingdom}
\affiliation{National Quantum Computing Centre, Didcot, OX11 0QX, U.K.}

\begin{abstract}
\textbf{
Universal blind quantum computing allows users with minimal quantum resources to delegate a
quantum computation to a remote quantum server, while keeping intrinsically hidden input, algorithm, and outcome. State-of-art experimental demonstrations of such a protocol have only involved one client. However, an increasing number of multi-party algorithms, e.g. federated machine learning, require the collaboration of multiple clients to carry out a given joint computation. In this work, we propose and experimentally demonstrate a lightweight multi-client blind quantum computation protocol based on a novel linear quantum network configuration (Qline). Our protocol originality resides in three main strengths: scalability, since we eliminate the need for each client to have its own trusted source or measurement device, low-loss, by optimizing the orchestration of classical communication between each client and server through fast classical electronic control, and compatibility with distributed architectures while remaining intact even against correlated attacks of server nodes and malicious clients.
}
\end{abstract}

\maketitle

\section{Introduction}

Despite the increasing technological progress in the manipulation of many-qubit systems \cite{bao2023very,chow2021ibm,arute2019quantum}, providing quantum computing as a service for end-users poses several challenges, including scalability, privacy, and integrity. Indeed, in order to achieve true quantum advantage from emerging devices, they must scale up beyond the current monolithically designed noisy intermediate-scale quantum regime. As of today, the only viable solution being pursued by all qubit platforms is modularity and interconnected architecture, where photonic links are considered the best option. Moreover, it is also clear that quantum machines need to be integrated into cloud services or data centers, allowing multiple clients to connect locally or globally to access these devices.
In such a context, the issue of keeping the computation and data protected from malicious parties will be a key challenge for such large-scale adaptation.  Notably, photonic links to quantum servers enable the capability of achieving informational security for delegated computing, known as blind quantum computing (BQC), which is not achievable using only classical communication between client and server \cite{aaronson2019complexity}.
Such a protocol builds on the measurement-based model for quantum computation \cite{raussendorf2001one, raussendorf2009measurement} that exploits mid-circuit measurements for teleportation-based quantum computing on encrypted quantum states sent to a remote server via a quantum link \cite{broadbent2009universal,leichtle2021verifying}. 
Over the last decade, many BQC protocols have been proposed \cite{liu2023public,kapourniotis2022framework,ma2022qenclave,gheorghiu2017rigidity,aharonov2017interactive,gheorghiu2015robustness,perez2015iterated,hajduvsek2015device,hayashi2015verifiable,morimae2014verification,morimae2013blind,sueki2013ancilla,mantri2013optimal,reichardt2012classical,morimae2012blind,dunjko2012blind}, together with proof-of-concept experimental demonstrations in different settings \cite{drmota2023verifiable,sheng2018blind,huang2017experimental,greganti2016demonstration,takeuchi2016blind,marshall2016continuous,fisher2014quantum,barz2013experimental,barz2012demonstration}. However, the challenge of multi-client settings has been explored only theoretically due to the high resource requirements of the proposed protocols \cite{kapourniotis2023asymmetric,shan2021multi,qu2021secure,ciampi2020secure,kashefi2017multiparty}. 

Yet, a growing number of classical delegated computing tasks require that multiple clients collaborate to carry out a joint function, e.g. federated machine learning tasks \cite{konevcny2016federated,yang2019federated}. Notably, quantum counterparts of such algorithms have been proposed as well \cite{chen2021federated}, including a federated quantum machine learning protocol based on BQC \cite{li2021quantum}.

\indent In this paper, we propose a modular lightweight distributed architecture for multi-client BQC based on the recently proposed Qline quantum network link configuration \cite{doosti2023establishing}, that enables scalable client insertion. With such an architecture in mind, we present a tailor-made multi-party blind quantum computing protocol such that the clients in the Qline only own trusted single qubit rotation devices, while the overall protocol still provides privacy for the joint computation performed by several users on the cloud. 
In the Qline network architecture, the quantum resource is first generated by a potentially untrusted server, then distributed to the clients, such that each client can apply arbitrary single-qubit operations on the incoming qubits and, at the end of the line, measured by a second again potentially untrusted server.
An analogous architecture was first introduced in \cite{hillery1999quantum} for quantum-assisted secret sharing, and later used for various tasks such as quantum key distribution or secure computation \cite{clementi2017classical,grice2015two,schmid2005experimental}. The main advantages of such a structure reside in the possibility to integrate it easily into larger-scale networks, its compatibility with key establishment protocols \cite{doosti2023establishing}, and its low hardware complexity.
In order to simplify the resource requirements of the multi-client BQC protocol in \cite{kashefi2017multiparty,kapourniotis2023asymmetric}, we show that, within such an architecture, it is enough that each client in the Qline adds a new layer of encryption to the flying qubits that will be used as the common key for their later private joint computation on the remote server. 
Such collaboration may be typical, for instance, of privacy-preserving machine learning algorithms where each client's input data and parameters related to the algorithm should remain private to all parties.
To implement it, we employ a fibered photonic platform equipped with genuine measurement adaptivity to enable deterministic computation \cite{raussendorf2001one}.
Within this setup, we are able to show the blindness and the correctness of the protocol, in both cases where the function to be computed has a classical or a quantum output. 
Our experimental proof-of-concept demonstrates a two-client scenario that can be easily extended to larger and more complex quantum networks featuring any number of clients at arbitrary distances.

\begin{figure}[t]
    \includegraphics[width=\columnwidth]{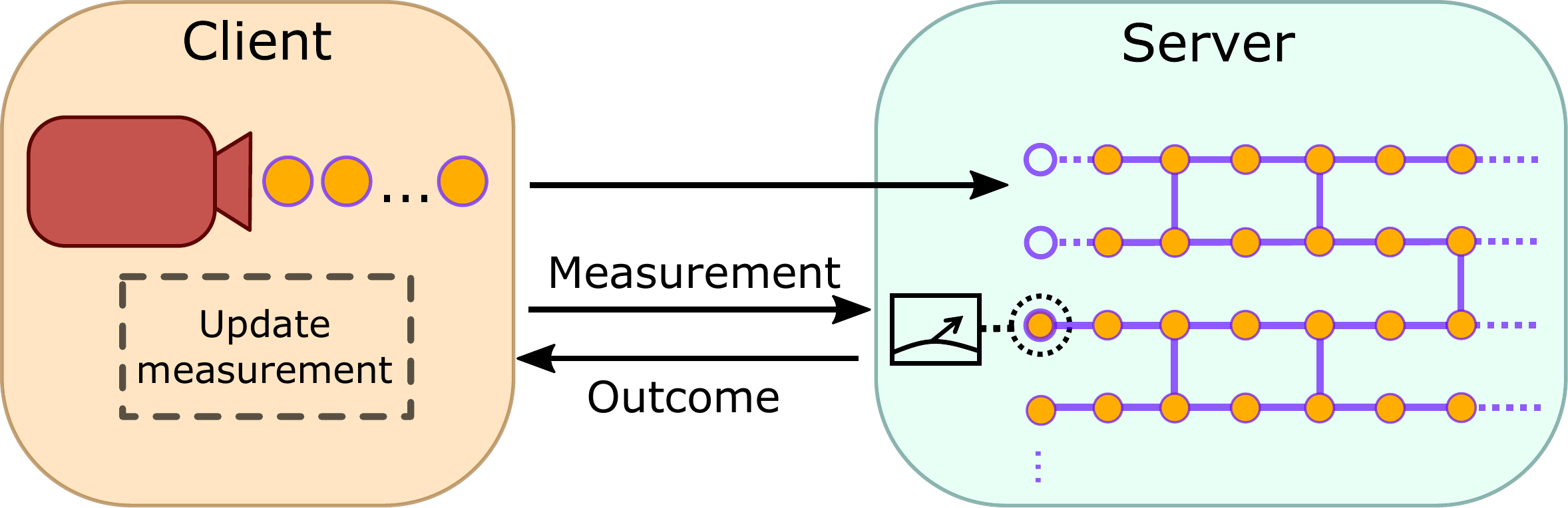}
    \caption{\textbf{Conceptual scheme of BQC.} In the BQC protocol, a client randomly prepares $n$ qubits and sends them to a quantum server. The server uses the qubits to form a resource state and, at each of the $\mathcal{O}(n)$ rounds of the computation, it measures one qubit in the measurement basis given by the client. Then, it gives back the measurement outcome to the client, who decides accordingly on the next measurement basis. 
    Figure inspired by \cite{fitzsimons2017private}.
    The details of BQC are presented in Protocol~S4 of the Supplementary Information.}
    \label{fig:scheme_bqc}
\end{figure}

\section{A multi-client BQC protocol}

In this section, we describe in detail the protocol proposed and successfully implemented in this work. 
It is built on the theoretical premises in \cite{kapourniotis2023asymmetric, kashefi2017multiparty}, and tailored to a Qline architecture \cite{doosti2023establishing}.  
Differently from the single client original protocol of \cite{broadbent2009universal} depicted in Fig.~\ref{fig:scheme_bqc}, the results of \cite{kapourniotis2023asymmetric, kashefi2017multiparty} enable multi-client BQC by exploiting secure multi-party computation (SMPC), whose aim is to allow several users to collaboratively compute a joint function on their private data. The classical SMPC functionality enables coordination of the parties in a delegated quantum computing task, such that the full computation details are blind not only to the server, but also to potentially dishonest clients that collude with it. However, implementing such functionality would need additional rounds of classical communication among the clients and server, during which it would be unfeasible to coherently store the quantum state. Therefore, to optimize the storage time, in our implementation, we substitute classical SMPC with a trusted third party (TTP) that mediates the communication between the clients and the server reducing the number of rounds, while the blindness of the protocol is still proven against any strict subset of colluding malicious adversaries. In this way, the quantum state needs to be stored for significantly shorter times than if using full classical SMPC, thus enabling our first proof-of-concept experimental demonstration of a two-client BQC.
To motivate our experimental design, this section is divided into two parts: the first one is devoted to the description of a two-client example, which we implemented experimentally demonstrating the key building blocks that are required for a fully scalable solution. In the second part, we present the extension to the $n$ clients case and universal quantum computing resources.

\subsection{The two-client protocol}
\begin{figure}[t]
    \includegraphics[width=\columnwidth]{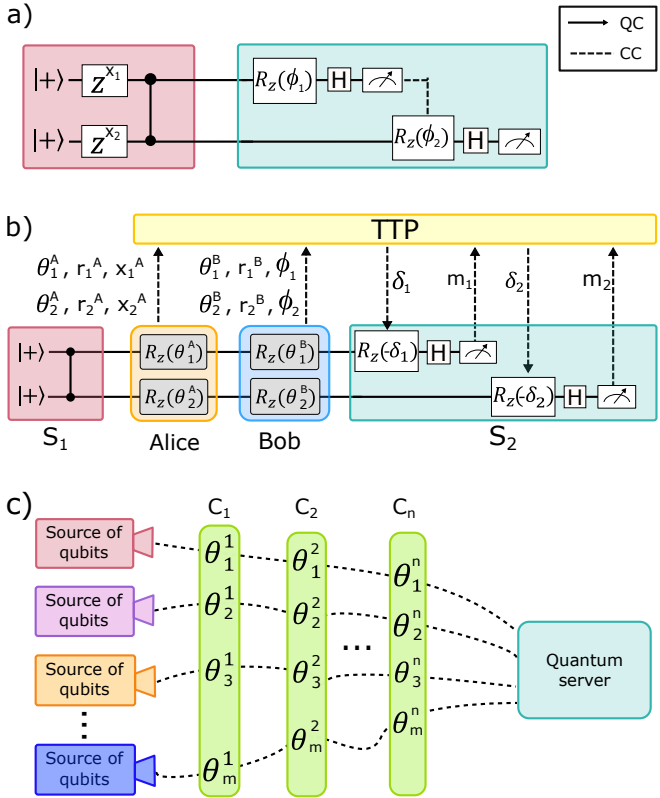}
    \caption{
    \textbf{Conceptual scheme of the two-client BQC and distributed quantum computing over a Qline architecture.
    a)} The desired joint quantum circuit computation where $x_1$ and $x_2$ are Alice's private input data and $\phi_1$ and $\phi_2$ the private angles of Bob's algorithm.
    \textbf{b)} The same computation of the circuit presented in \textbf{a)} is encrypted to preserve the privacy of each party's information. In our two-client BQC protocol, a quantum channel connects a source of bipartite quantum states to two clients disposed along the Qline. Each client chooses their secret parameters $\theta_i^j,~r_i^j,~x_i,~\phi_i$, for $i=1,~2$ and $j=A,~B$, and applies $z$-rotations to both qubits. All secret parameters and measurement outcomes pass through a TTP to compute the transformed measurement bases $\delta_i$ and the corrected outcomes.
    At the end of the line, the quantum state is sent to the server $S_2$ to carry out the desired computation, which is carried out through two rounds of classical communication between clients, TTP, and the server $S_2$.
    \textbf{c)} Generalization of our architecture to $m$ Qlines with $n$ clients distributed along them. In a Qline, a quantum state source distributes single qubits and each client applies random rotations to all qubits. At the end of the line, a powerful quantum server employs the received qubits to generate the resource state for the computation and calculate a joint function.
    }
    \label{fig:conceptual_schemes}
\end{figure}

 Consider Alice and Bob wish to run a joint computation on a remote server. Alice has private classical data $x_1$ and $x_2$ while Bob has private gate parameters $\phi_1$ and $\phi_2$, and the target joint circuit is
 \begin{equation*}
\left(M^{X}\otimes M^{X} \right)  \left(R_z(\phi_1)\otimes R_z(\phi_2)\right)  ~ \mathsf{CZ}_{12}  ~ \left(Z^{x_1} \otimes Z^{x_2} \right) \left( \ket{+} \otimes \ket{+} \right)
\end{equation*}

  as shown in Fig.~\ref{fig:conceptual_schemes}a. This is a typical building block of any large-scale privacy-preserving QML, such as the one proposed in \cite{chen2021federated}. In what follows we demonstrate the steps to make the above joint computation both distributed and secure as shown in Fig.~\ref{fig:conceptual_schemes}b.

\vspace{1mm} 

\textbf{State preparation.}
A source of maximally entangled bipartite states, $S_1$, distributes two-qubit states along two quantum channels, of the form:
\begin{equation}
    \ket{\psi} = \frac{1}{2} \left( \ket{00} + \ket{01} + \ket{10} - \ket{11} \right)
\label{eq:generated_state}
\end{equation}
Alice receives the two qubits, and applies single-qubit $z$-rotations of angles $\theta_i^{A}$ to them, randomly chosen in the set $\mathcal{A} = \{ 0,~ \pi/4,~ 2\pi/4, ... ~7\pi/4 \}$.
This will hide (via quantum one-time padding) her classical input data which would be encoded on these qubits via $Z^{x_i}$ operations. Moreover, she  chooses two random bits $r_1^A,~ r_2^A$ that will later hide the outcome of the computation.  She communicates her secret parameters to the TTP.
She then sends her two encrypted qubits to the second client, Bob, who applies further random $\theta_1^B,~ \theta_2^B$ $z$-rotations, again chosen from the set $\mathcal{A}$ to one-time pad his private algorithm parameter $\phi_1,~ \phi_2$. Bob also chooses two random bits $r_1^B,~ r_2^B$ for the encryption of the output as well. He then communicates his secret parameters to the TTP.
From now on, we will use the following definitions: $\theta_i = \theta_i^A + \theta_i^B$ and $r_i = r_i^A \oplus r_i^B$. The resulting quantum state at this stage is the following:
\begin{equation}
\ket{\psi} = \frac{1}{2} \left( \ket{00} + e^{i\theta_2}\ket{01} + e^{i\theta_1}\ket{10} - e^{i(\theta_1 + \theta_2)}\ket{11} \right)
\label{eq:rotated state}
\end{equation}
This state is then sent to server $S_2$.

\vspace{1mm} 

\textbf{Interaction and measurement stage.} 
From now on, the clients and $S_2$ only communicate classically, through the TTP. The protocol requires two rounds, one for each qubit to be measured. The blind measurement angle $\delta_i$ at the $i-$th round, for $i=1,~2$, is computed by the TTP according to the formula:

\begin{equation}
\delta_i = \theta_i  + x_i\pi + (-1)^{m_{(i-1)}^{true}}\phi_i + r_i \pi
\label{eq:deltas}
\end{equation}
\\
where $m_0^{true} = 0$ and $m_1^{true} = m_1 \oplus r_1$. Analogously to the first measurement outcome, the outcome of the second measurement is decrypted according to the formula: $m_2^{true} = m_2 \oplus r_2$ before giving it back to the clients.
A slight change in the protocol is needed in the case where a quantum function is computed, i.e. the output qubit must be prepared by the clients in the state $\ket{+}$ \cite{broadbent2009universal}.
The correctness of the protocol is straightforward and can be directly obtained from \cite{broadbent2009universal,kashefi2017multiparty,kapourniotis2022framework}. However, the security proof is more subtle compared to previous works that were based on trusted state preparation for each client. We first present the generalization of our protocol and then provide the full proof of security that is applicable to this special case as well.

\subsection{Generalization to the multi-client scenario}

In this section, we generalize our protocol to a scenario where $n$ clients want to perform a joint computation on a possibly larger resource state.
Blindness should be guaranteed for any single honest client.
We consider the target computation to be defined as a measurement pattern \cite{raussendorf2001one} by the measurement angles $(\phi_v)_{v\in V}$, where $v\in V$ ranges over all qubits in the resource graph state. These angles can be fixed and publicly known, or jointly input by any subset of clients. In the latter case, blindness holds for the measurement angles as well.
The input qubits $I\subset V$ are partitioned into sets $(I_j)_{j\in \{1,\dots,n\}}$, where $I_j$ belongs to the $j$-th client who has the bit string $x_j \in \{0,1\}^{|I_j|}$ as input.
To keep each client's input blind, it is required that each qubit in the resource state travels once along the Qline and accumulates random rotations by all clients, as depicted in Fig.~\ref{fig:conceptual_schemes}c.
In this way, a resource state on $m = |V|$ qubits would require $m$ Qlines.
However, this process may be (partially) parallelized, in the sense that multiple qubits can be sent along the Qline at once -- as long as the clients can perform the necessary rotations in parallel as well.
After the server received the qubits that have passed through the Qline, it follows a standard execution of the BQC protocol (Protocol~S4 of the Supplementary Information. As all communication from this point forward is entirely classical, the TTP governs the instructions to the server for the remainder of the protocol.
A detailed description of the full multi-client protocol on scalable resource states is given in Protocol~\ref{prot:mcbqc}.

\begin{protocol}
    \caption{Multi-Client Blind Quantum Computation}
    \label{prot:mcbqc}
    \DontPrintSemicolon
    \SetKwInput{Public}{Public Information}
    \SetKwInput{Inputs}{Inputs}
    \SetKwInput{Protocol}{Protocol}
    \Public{
        \begin{itemize}
            \item A graph $G = (V, E, I, O)$ with input and output vertices $I$ and $O$, respectively.
            \item A partition $(I_j)_{j\in\{1,\dots,n\}}$ of the input vertices, where $I_j$ belongs to client $j$.
            \item A partial order $\preceq$ on the set $V$ of vertices.
        \end{itemize}
    }
    \Inputs{
        \begin{itemize}
            \item Client $j$ has a classical bit string $x_j \in \{0,1\}^{|I_j|}$.
            \item The $n$ clients collaboratively input a set of angles $(\phi_v)_{v\in V}$, and a flow $f$ on $G$ compatible with $\preceq$.
            \item The orchestrator and the server have no inputs.
        \end{itemize}
    }
    \Protocol{
        \begin{enumerate}
            \item The $n$ clients send all of their inputs to the orchestrator.
            \item The orchestrator and the server perform the BQC Protocol~S4. For every $\ket{+_\theta}$-state that the orchestrator would need to send to the server, they instead perform the following:
                \begin{enumerate}
                    \item The server prepares a $\ket{+}$-state.
                    \item All parties perform one execution of the Collaborative Remote State Rotation Protocol~S5, where the server uses the $\ket{+}$-state as an input, and the orchestrator inputs the angle $\theta$.
                    \item The server uses the output state in the BQC Protocol~S4.
                \end{enumerate}
            \item The orchestrator distributes the classical output among the clients. The server sends the output qubits in $O$ to the designated clients, with the orchestrator providing the decryption keys.
        \end{enumerate}
    }
\end{protocol}

Note that the orchestrator from Protocol~\ref{prot:mcbqc} must be trusted, but is an entirely classical party. 
To remove all trust assumptions on this party, it can be replaced with any composably secure classical SMPC protocol which is performed by the clients and the server.
Moreover, much of the calculations that the orchestrator needs to perform, including the sampling of random coins and the evaluation of the formulae for the corrected measurement angles, can be done in a classical pre-processing step. The computation during the quantum phase then boils down to the choice of one of two possible measurement angles based on the previously reported measurement outcomes.
Finally, it is worth mentioning that the requirement that every client has access to each Qline can be weakened when accepting stronger trust assumptions. Blindness holds as long as there exists at least one honest client along each Qline. Therefore, even if not every client participates at each Qline, blindness is still guaranteed if we restrict the adversarial patterns to not corrupt all clients along any Qline at once.

In the concrete case of our experiment, the graph $G$ is a two-qubit cluster state, that is, $V=\{1,2\}$. Both qubits are Alice's input qubits, so $I_1 = I = V$, while Bob has no input qubits ($I_2 = \emptyset$) but chooses the measurement angles $(\phi_1, \phi_2)$.
We consider two different computations: in the first case, we interpret both measurement outcomes as the classical output of the computation, while in the second case, we consider the second qubit to be the quantum output of the computation, hence $O = \{2\}$.

\subsection{Security}

In our protocol, the clients' quantum abilities are restricted to receiving single-qubit states, applying a random $z$-rotation to it, and forwarding it to the next client or server. From the perspective of an honest client, this behavior is exactly captured by the \emph{Remote State Rotation} (RSR) functionality introduced in~\cite{ma2022qenclave}. The latter showed that RSR can indeed be used in the context of the BQC protocol to delegate a universal quantum computation with perfect blindness. This immediately implies the security of the proposed protocol with a single honest client and an untrusted server, as the instructions of the protocols exactly coincide.
By a similar argument, security can be shown for the two-client and multi-client generalizations. In the spirit of the security proof in~\cite{kapourniotis2023asymmetric}, the sequential rotations of a single qubit by all clients along the Qline can be seen as a collaborative version of RSR where the role of the RSR-client is now taken by an entirely classical trusted virtual party, the \emph{orchestrator}. The entire execution of the protocol then becomes one run of the BQC protocol between the orchestrator and the server. Finally, in real-world implementations, the orchestrator can be replaced by a classical SMPC protocol. 
Security follows by composition of the above-mentioned building blocks which have all been proven to be composably secure in the Abstract Cryptography framework~\cite{maurer2011abstract}. 
Further details about the security analysis of the full protocol, including a formal security proof, are provided in Sec.~S1 of the Supplementary Information.

\begin{figure*}[t]
    \includegraphics[width=0.8\textwidth]{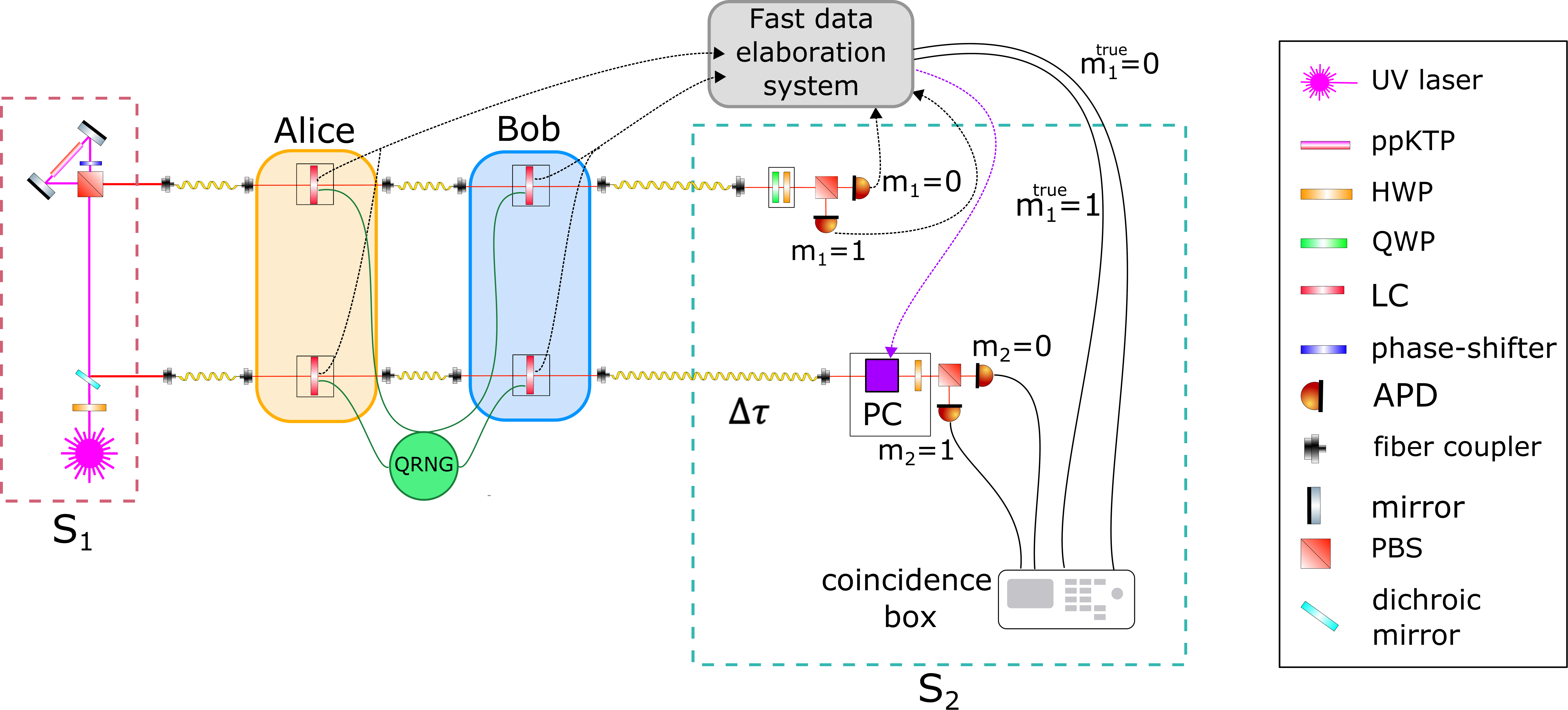}
    \caption{\textbf{Experimental apparatus.} The state defined in Eq.~\eqref{eq:generated_state} is generated through a Sagnac-based source of entangled photons, where horizontal polarization ($\ket{H}$) encodes the state $\ket{0}$ while vertical polarization ($\ket{V}$) encodes the state $\ket{1}$. The photons are first sent to Alice and Bob who performs single-qubit rotations by means of liquid crystals (LC). To make the clients' choices random, the clients' secret parameters are chosen by means of a quantum random number generator (QRNG).
    Then, the two photons are sent to Server 2 (S2). While the second photon is delayed through a $\approx 65$ m-long single-mode fiber, the first photon is measured, by using a sequence of a quarter-wave plate (QWP), a half-wave plate (HWP), and a polarizing beam splitter (PBS). The second measurement station, instead, is composed of a Pockels cell (PC), a HWP, and a PBS.
    A fast data elaboration system, constituted by a computer and a fast electronic circuit, embodies the TTP. The two detectors of the first measurement station are linked to such a system, that directly activates the PC with a suitable high-voltage (HV) pulse according to the desired second measurement basis. All outcomes are collected through a coincidence box that records as two-fold coincidences all events occurring in a given time window.} 
    \label{fig:exp_setup}
\end{figure*}

\section{Experimental apparatus}\label{sec:setup}
In Fig.~\ref{fig:exp_setup} we describe the experimental apparatus we employ to implement the two-client protocol. A detailed time scheme of the protocol is reported in Sec.~S2 of the Supplementary Information.

\vspace{1mm} 
\textbf{Photon source.} A Sagnac-based source of polarization-entangled photons, i.e. server $S_1$, generates pairs of photons in the state defined in Eq.~\eqref{eq:generated_state}, where we encode the computational basis vector $\ket{0}$ in the photons' horizontal polarization ($\ket{H}$), and $\ket{1}$ in the vertical one ($\ket{V}$). 
The photon pairs are sent to the clients who apply their random rotations. The resulting state after these transformations is defined in Eq.~\eqref{eq:rotated state}.

\vspace{1mm} 
\textbf{Clients' preparation.}
At each run of the protocol, we set all random parameters through an ID Quantique quantum random number generator (QRNG).  Both clients use liquid crystals (LCs) to apply their rotations, which are set in this preparation stage. The TTP is made up of a computer linked to a fast electronic circuit and stores all clients' parameters. With such information, the TTP pre-computes the measurement angle $\delta_1$, and the first measurement station is set accordingly. Moreover, the TTP also pre-computes the two possible values for $\delta_2$, considering that the first outcome is still unknown, namely $ \delta_2^{\pm} = \theta_{2} + x_2\pi + r_{2}\pi \pm \phi_2 $, to speed up the measurement step of the protocol.

\vspace{1mm} 
\textbf{Measurement.} 
The two photons are sent to server $S_2$ of Fig.~\ref{fig:exp_setup} where measurements of the form $M(\delta) = \cos{(\delta)}\sigma_x + \sin{(\delta)}\sigma_y$ are performed.
The first one is made up of a quarter-wave plate (QWP), a half-wave plate (HWP), and a polarizing beam splitter (PBS).
The two single-photon avalanche photodiodes (APD) of the first measurement station are connected to a fast electronic circuit that selects $ \delta_2^{+}$ or $ \delta_2^{-}$, according to the corrected outcome of the first measurement, $m_1^{true} = m_1 \oplus r_1$.
In the second measurement station, we substitute the QWP with a Pockels cell (PC), i.e. a fast electro-optical modulator that performs the identity when no voltage is applied, while applying a phase shift between orthogonal polarizations when a voltage is applied.
The second photon is delayed with respect to the first one by using a $\approx 65$ m single-mode fiber to enable feed-forward in the second measurement station.
Finally, the second outcome is corrected according to $m_2^{true} = m_2 \oplus r_2$.
Further details about the feed-forward system can be found in Sec.~S4 of the Supplementary Information. All events are collected through a coincidence box that records as two-fold coincidences all detector clicks occurring in a given time window.

\section{Results}
\begin{figure}[httb]
    \includegraphics[width=\columnwidth]{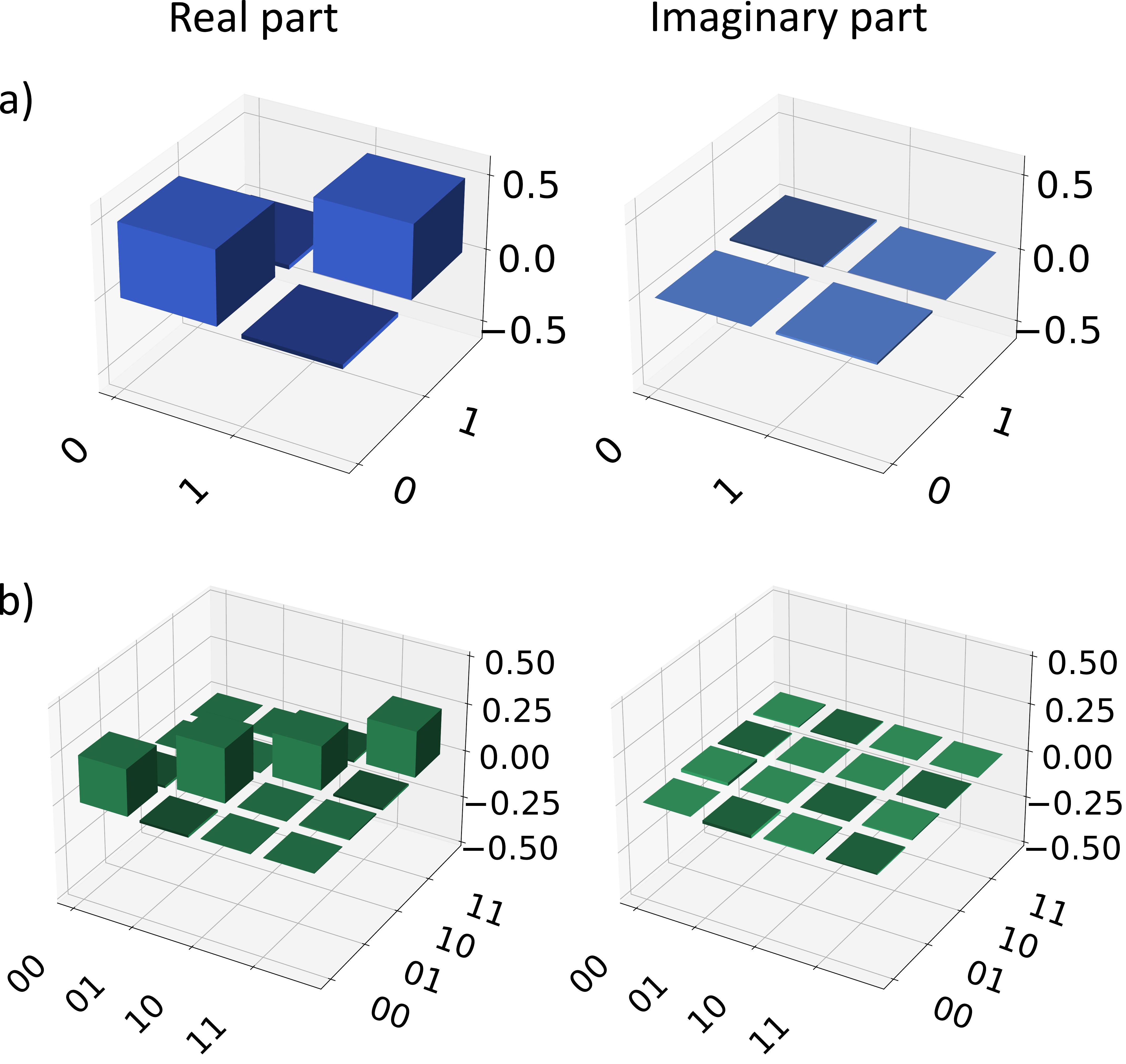}
    \caption{\textbf{Demonstration of blindness. a)}  
     Density matrix of the second qubit averaged over all possible $\theta_1^A$ and $\theta_1^B$ configurations.  \textbf{b)} 
    Density matrix of the two-qubit initial state, averaged over all possible values of $\theta_1^A$ and $\theta_2^B$.}
    \label{fig:tomo blind qubits}
\end{figure}

\textbf{Blindness of the protocol.} To show that the server cannot gain any information about the outcome of the computation, we suppose that the clients want to compute a given quantum function whose outcome is represented by the second qubit.  We repeated the experiment for both qubits of the cluster state, but we show in the main text only the resulting density matrix for the second qubit. Details about the blindness of the first qubit can be found in Sec.~S5 of the Supplementary Information. We demonstrate blindness of the second qubit by keeping the measurement angle $\delta_1 = \pi$ fixed and by averaging over all density matrices resulting in the output qubit for different initial rotation angles $\theta_1^{j}$, where $j = A,B$, namely $64$ combinations.
The density matrix in Fig.~\ref{fig:tomo blind qubits}a shows the resulting quantum state, which has a fidelity with a single-qubit completely mixed state amounting to $F_2 = 0.99870 \pm 0.00003 $, while the measured Von Neumann entropy is $S_2 = 0.9963 \pm  0.0001 $, to be compared with the expected value of 1 for a completely mixed single-qubit state.
Furthermore, we demonstrate the blindness of the whole initial two-qubit cluster state, by averaging over all density matrices corresponding to $64$ combinations of the initial $z$-rotations, for parameters $\theta_1^A$ and $\theta_2^B$, while keeping $\theta_1^B, ~\theta_2^A = 0$. We stress that these combinations are enough to demonstrate blindness, as the random rotations on each qubit still take all possible values in the set $\mathcal{A}$.
For the two-qubit state, whose density matrix is shown in Fig.~\ref{fig:tomo blind qubits}b, we estimated a fidelity $F = 0.99433 \pm 0.00003 $ with the completely mixed state and with a Von Neumann entropy of $S = 1.9836 \pm 0.0001 $, to be compared with the expected value of 2 for a completely mixed two-qubit state. All density matrices are retrieved from raw experimental data through quantum state tomography \cite{nielsen2002quantum}.
Further details about the blindness of the second qubit and the full initial state are reported in Secs.~S6-S7 of the Supplementary Information.

\begin{figure}
    \centering
    \includegraphics[width=\columnwidth]{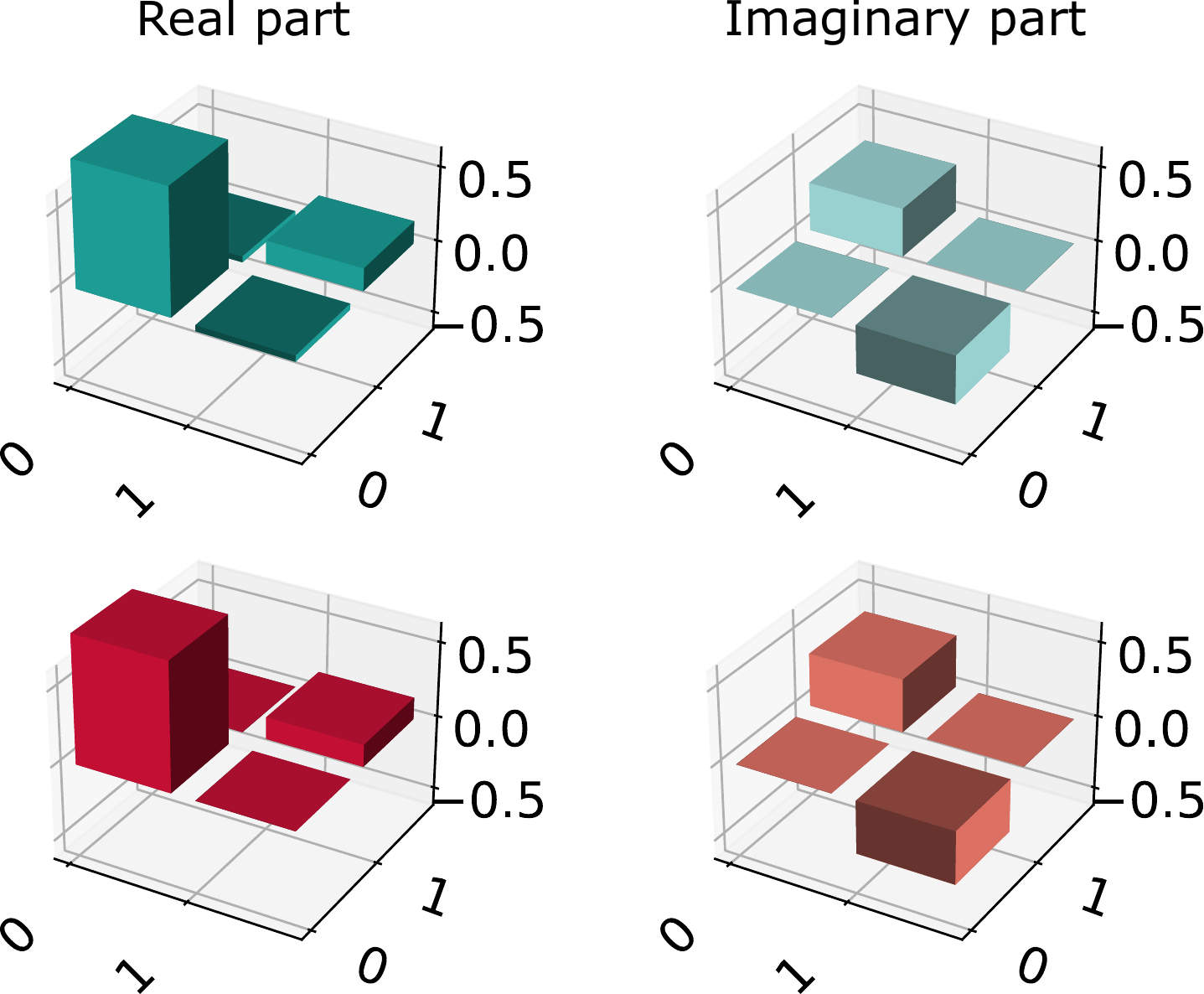}
    \caption{\textbf{Computation of a quantum function.} In this figure, we show the results of the computation of a quantum function. Bob chose $\phi_1 = \pi/4$, while the clients' random parameters are  $\theta_1^{A} = \pi/2$ and $\theta_1^{B} = \pi/4$. Both input bits $x_1,~x_2$ are set to $0$. The first qubit is thus measured in the basis $\delta_1 = \pi$. The experimental density matrix is shown in light blue, while the theoretical one is shown in red.}
    \label{fig:algoritmi_quantum}
\end{figure}

\begin{figure}[httb]
    \includegraphics[width=\columnwidth]{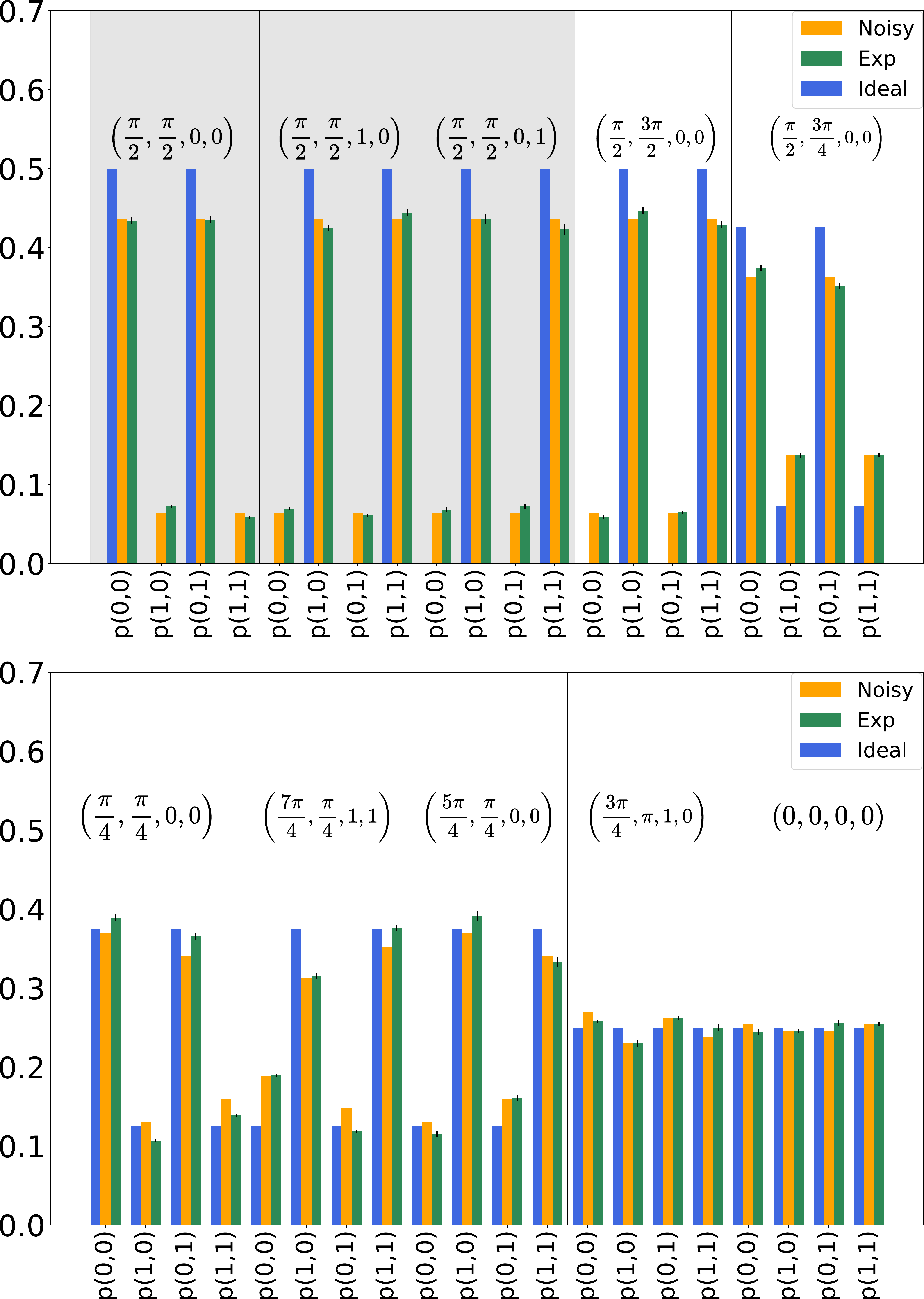}
    \caption{\textbf{Computation of a classical function}. In this bar plot, we show ten different measurement angles and Alice's input combinations $\left( \phi_1,~\phi_2,~ x_1,~x_2\right)$. In the grey region, we kept the algorithm fixed while changing the input data, to show the changes in both the expected and experimental distributions. In the white region, instead, we changed both algorithms and input data. 
    The uncertainties on the experimental frequencies were obtained assuming Poissonian statistics, and the black bars correspond to one standard deviation. The eventual absence of black bars means that the uncertainty was too small to be visible in this plot.}
    \label{fig:colorprobs_ff}
\end{figure}

\vspace{1mm} 
\textbf{Correctness of the protocol.} Let us now consider the scenario where the clients want to compute a quantum function. In this case, we take the second qubit as the outcome of the computation, by preparing it in the state $\ket{+}$. We perform the computation $\phi_1 = \pi/4$, with input bits $x_1,~x_2$ set to $0$. We set the clients' secret parameters as $\theta_1^{A} = \pi/2$, $\theta_1^{B} = \pi/4$ and $r_1^{A} = r_1^{B} = 0$.
We show our results in Fig.~\ref{fig:algoritmi_quantum}. 
The estimated fidelity with the ideal state amounts to $F_{\pi/4} = 0.972 \pm 0.003$.
In Sec.~S5-S6 of the Supplementary Information, we show instances of single-qubit quantum state tomographies of the two qubits and their fidelity with respect to the theoretical expectations.
We then demonstrate the correctness of the computation for  classical functions. The algorithm performed over input data $x_1, ~x_2$ is characterized by the two true measurement angles $\phi_1,~\phi_2$. Choosing $x_1$ or $x_2$ equal to 1 has simply the effect of inverting the minima and the maxima of the distributions. In Fig.~\ref{fig:colorprobs_ff}, we show ten different probability distributions obtained by trying ten different combinations of the algorithm parameters $\left(  \phi_1,~\phi_2,~x_1,~x_2 \right)$, and the comparison with the ideal and the noisy model case. 
All the obtained results are in good agreement with our noisy model and follow qualitatively the ideal expectations. Small deviations from the expected values are mainly due to the visibility of the quantum state at the end of the Qline and to imperfections coming from the non-ideal electro-optical modulators employed, i.e. the LCs and the PC (further details about our noisy model are in Sec.~S8 of the Supplementary Information).

\section{Discussion}
In this work, we proposed a multi-client version of the BQC protocol \cite{broadbent2009universal} and experimentally demonstrated it in a two-client setting.
We first simplified the protocol described in \cite{kapourniotis2023asymmetric,kashefi2017multiparty} to tailor it to the photonic Qline network introduced in \cite{doosti2023establishing}. To this end, we studied a photonic platform equipped with a source of polarization-entangled photon pairs, an active feed-forward system, and a fiber-based structure to connect the involved parties. In our scheme, the clients only need to apply single-qubit rotations.
Within this setup, we computed the outcomes of ten different classical functions, by changing the input data and the algorithm, and compared the results with a noisy model compatible with our experimental conditions. Also, we demonstrated the correctness of the protocol when the function to be computed has a quantum output.
Finally, we showed that the server cannot gain any information about the inputs of the clients or the outcome of the computation.

\vspace{1mm} 
\textbf{Scalability.}
Our proof-of-concept demonstration can represent a step forward toward the realization of a scalable and secure quantum cloud access infrastructure with multiple clients.
Indeed, in a real-world protocol, the necessary classical communication as part of the SMPC in between the measurements would considerably increase the time latency, in particular, if run over a slow network, such as the internet. Therefore, to make the experimental realization of the proposed protocol feasible, we replaced the SMPC with a fast electronic data elaboration circuit to both reduce the communication rounds and compensate for the absence of a quantum memory.
This allowed us to reduce the delay between measurements to a manageable level. 
Scaling up the size of the computation would require more qubits and their communication from the photon source past all clients to the server. In principle, this could be realized in two ways that can eventually be combined. First, the qubits could be sent all at once, which would require the clients to be able to apply rotations to multiple states at once. Alternatively, the qubits could be sent sequentially, one at a time. However, this second option would require the ability to store or delay them until the clients were able to adjust their rotation gates since every qubit is rotated by a different angle. In our demonstration, we opted for the first option as it represented the optimal way to minimize time latency and, consequently, photon loss. 
Moreover, the choice of adopting an optimized feed-forward system for measurement adaptivity in our setup is not only crucial to ensure blind and deterministic computation, but also to scale up the protocol. Indeed, post-selection schemes would require an exponentially growing number of measurements depending on the dimension of the quantum system, which would affect significantly the possibility of applications to larger quantum systems.

\vspace{1mm} 
\textbf{Future work.}
While our implementation guarantees the privacy of the inputs provided by the clients, the outcome of the computation is not verified. We leave the addition of verification to the proposed protocol as future work. One possible path towards verification with Qline architecture might be the employment of the novel \emph{dummyless} testing technique from \cite{kapourniotis2023asymmetric}. However, as of now, the question whether states prepared by rotation-only clients are sufficient for verification remains open, as \cite{ma2022qenclave} only showed that this kind of states are enough to achieve blindness.\\

We believe that this work has insightful implications both from a theoretical and an experimental point of view. From a theoretical perspective, it constitutes a strong encouragement toward the development of collaborative computational algorithms over distributed quantum networks, as well as investigations about their verification. From an experimental point of view, instead, it represents a step forward toward the applications of photonic linear quantum networks as building blocks for more complex networks, toward the realization of a large and densely connected quantum cloud.

\section{Methods}
Photon pairs are generated in a parametric down-conversion source, composed of a 25-mm-thick  periodically-poled Potassium Titanyl Phosphate (ppKTP) crystal inside a Sagnac interferometer. The source uses a Toptica continuous-wave diode laser with a wavelength equal to $405~$nm. Both photons are generated at a wavelength equal to $810~$nm. To test the quality of the bipartite state generated by $S_1$, we perform a CHSH Bell test \cite{clauser1969proposed} and obtain a Bell parameter equal to $2.752 \pm 0.006$.
The generated photons are filtered in wavelength and spatial mode by using, respectively, narrow-band filters and single-mode fibers. The PC is a LiNbO$_3$ crystal made by the Shangai Institute of Ceramics having a rise-time equal to $90$~ns. A fast electronic circuit transforms signals coming from the detectors of the first measurement station into high-voltage calibrated pulses, needed to activate the PC. 
The amount of delay on the second photon was evaluated considering the response time of the detectors, the speed of the signal transmission through a single-mode fiber, whose refraction index is $\approx 1.45$, and the activation time of the PC. Therefore, we used a $\approx 65~$m long single-mode fiber that allows a delay of $\approx 320~$ns of the second photon with respect to the first. The voltages applied to the PC to insert a phase shift equal to $\pi/4, ~\pi/2,~ 3\pi/4$ were, respectively, $V_{\pi/4} = 650~$V,$~ V_{\pi/2} = 850~$V,$~ V_{3\pi/4} = 1100~$V. 
Further details about the feed-forward system are given in Sec.~S4 of the Supplementary Information. Our experiment is performed shot-by-shot, namely, each event of our data takings is characterized by a different randomly chosen set of initial parameters $\theta^j_i, r_i^j$, for $i=1,~2$ and $j=A,~B$, while the algorithm ($\phi_i,~x_i$, for $i=1,~2$) is kept fixed for each data taking.

\section*{Author contributions}
B. P., D. L., L. L., G. C., N. S., M. K., F. S., and E. K. contributed to developing the ideas and discussions explored in the paper.  B. P., D. L., M. K., and E. K. contributed to the design of the protocols. B. P., L. L., G. C., G. M., N. S., and F. S. contributed to the experimental implementation.  All authors contributed equally to the writing, editing, and final revision of the paper.

\section*{Acknowledgements}
This work was supported by the European Union’s Horizon 2020 research and innovation program through the FET project PHOQUSING (``PHOtonic Quantum SamplING machine'' – Grant Agreement No.
899544) and by the ANR research grant ANR-21-CE47-0014 (SecNISQ).

\end{document}


\title{Supplementary Information}

\author{Beatrice Polacchi}
\affiliation{Dipartimento di Fisica - Sapienza Universit\`{a} di Roma, P.le Aldo Moro 5, I-00185 Roma, Italy}

\author{Dominik Leichtle}
\affiliation{Laboratoire d’Informatique de Paris 6, CNRS, Sorbonne Université, 75005 Paris, France}

\author{Leonardo Limongi}
\affiliation{Dipartimento di Fisica - Sapienza Universit\`{a} di Roma, P.le Aldo Moro 5, I-00185 Roma, Italy}

\author{Gonzalo Carvacho}
\affiliation{Dipartimento di Fisica - Sapienza Universit\`{a} di Roma, P.le Aldo Moro 5, I-00185 Roma, Italy}

\author{Giorgio Milani}
\affiliation{Dipartimento di Fisica - Sapienza Universit\`{a} di Roma, P.le Aldo Moro 5, I-00185 Roma, Italy}

\author{Nicolò Spagnolo}
\affiliation{Dipartimento di Fisica - Sapienza Universit\`{a} di Roma, P.le Aldo Moro 5, I-00185 Roma, Italy}

\author{Marc Kaplan}
\email{kaplan@veriqloud.fr}
\affiliation{VeriQloud, 13 rue Victor Hugo, 92 120 Montrouge, France}

\author{Fabio Sciarrino}
\email{fabio.sciarrino@uniroma1.it}
\affiliation{Dipartimento di Fisica - Sapienza Universit\`{a} di Roma, P.le Aldo Moro 5, I-00185 Roma, Italy}

\author{Elham Kashefi}
\email{elham.kashefi@lip6.fr}
\affiliation{Laboratoire d’Informatique de Paris 6, CNRS, Sorbonne Université, 75005 Paris, France}
\affiliation{School of Informatics, University of Edinburgh, 10 Crichton Street, EH8 9AB Edinburgh, United Kingdom}

\maketitle

\tableofcontents

\section{Security Analysis}\label{sec:security_analysis}

In this section, we provide the details of the security proof for the full multi-client blind quantum computation protocol. The security analysis uses techniques and results from previous work on the Qline architecture~\cite{doosti2023establishing}, on quantum secure multi-party computing (QSMPC)~\cite{kapourniotis2023asymmetric}, and on blind delegated computing using trusted Remote State Rotation (RSR)~\cite{ma2022qenclave}.

\subsection{Abstract Cryptography}

The security proof is given in the Abstract Cryptography (AC) framework~\cite{maurer2011abstract}. The AC framework is used to define cryptographic protocols, and analyze their security. As opposed to stand-alone and game-based security models, it allows the secure composition of protocols within the framework. In this sense, if two or more protocols have been proven to be AC-secure, so is their sequential or parallel composition.

In the following, we refer to protocols, usually denoted by $\pi$, and resources, which we will denote by $\mathcal{R}$. Protocols can be used to construct new resources from other resources, denoted $\mathcal{R}_1 = \pi \mathcal{R}_0$. In order to define security for such constructions, one requires a sense of distance and equivalence in the space of resources. This is captured by the following notion of indistinguishability.

\begin{definition}[Indistinguishability of resources]
    Let $\mathcal{R}_0$, $\mathcal{R}_1$ be resources with matching input and output interfaces. They are called $\varepsilon$-(statistically)-indistinguishable, denoted $\mathcal{R}_0 \approx_\varepsilon \mathcal{R}_1$, if for all (unbounded) distinguishers $\mathcal{D}$ it holds that
    \begin{align*}
        \left| \Pr \left[ b=1 | b \gets \mathcal{D}\mathcal{R}_0 \right] -  \Pr \left[ b=1 | b \gets \mathcal{D}\mathcal{R}_1 \right] \right| \leq \varepsilon.
    \end{align*}
\end{definition}

Equipped with the pseudo-metric induced by the indistinguishability of resources, we can now proceed to define what it means to securely construct a resource, even in the presence of malicious parties, i.e., parties that deviate from the instructions specified by the protocol.

\begin{definition}[Secure construction of resources]
    Let $\pi$ be an $n$-party protocol, and let $P \subseteq \mathfrak{P}(\{1,\dots,N\})$. $\pi$ is then said to $\varepsilon$-securely construct resource $\mathcal{R}_1$ from resource $\mathcal{R}_0$ against adversarial patterns $P$ if the following two properties hold:
    \begin{enumerate}
        \item Correctness: $\pi \mathcal{R}_0 \approx_\varepsilon \mathcal{R}_1 \perp$, where $\perp$ is used to filter all malicious interfaces;
        \item Soundness: for all subsets $M \in P$ of colluding corrupted parties, there exists a simulator $\sigma_M$ such that $\pi_{M^c} \mathcal{R}_0 \approx_\epsilon \mathcal{R}_1 \sigma_M$.
    \end{enumerate}
\end{definition}

As mentioned previously, one power of the AC framework lays in the secure composability of resources. The following theorem \cite[Theorem~1]{maurer2011abstract} is the formalization of this claim.

\begin{theorem}[General composition of resources]\label{thm:ac-composition}
    Let $\mathcal{R}_0, \mathcal{R}_1, \mathcal{R}_2$ be resources, $\pi_0, \pi_1$ be protocols, and $\mathsf{id}$ be the identity protocol. Let $\circ$ denote the sequential composition, and $|$ denote the parallel composition of protocols and resources, respectively. The following statements then hold:
    \begin{enumerate}
        \item Sequential composition: $\pi_0 \mathcal{R}_0 \approx_{\varepsilon_0} \mathcal{R}_1$ and $\pi_1 \mathcal{R}_1 \approx_{\varepsilon_1} \mathcal{R}_2$ imply $(\pi_1 \circ \pi_0) \mathcal{R}_0 \approx_{\varepsilon_0 + \varepsilon_1} \mathcal{R}_2$;
        \item Context-insensitivity: $\pi_0 \mathcal{R}_0 \approx_{\varepsilon} \mathcal{R}_1$ implies\\ $(\pi_0 | \mathsf{id})(\mathcal{R}_0 | \mathcal{R}_2) \approx_{\varepsilon} (\mathcal{R}_1 | \mathcal{R}_2)$.
    \end{enumerate}
\end{theorem}

Note that the combination of sequential composition and context-insensitivity yields the security of parallel composition as well.

\subsection{Ideal functionalities}

In the AC framework, the desired, perfect behavior of protocols is captured by resources that we call ideal functionalities. We present the ideal functionalities of the resources relevant to the presented security proof in the following.

Resource~\ref{res:rsr} describes the \emph{Remote State Rotation} functionality, as defined in \cite[Definition~7]{ma2022qenclave}. Its purpose is to capture the quantum abilities of the clients which are limited to performing single-qubit rotations around the $z$-axis.

\begin{resource}
    \caption{Remote State Rotation}
    \label{res:rsr}
    \DontPrintSemicolon
    \SetKwInput{Inputs}{Inputs}
    \SetKwInput{Comp}{Computation by the resource}
    \Inputs{
        \begin{itemize}
            \item The client sends an angle $\theta \in \mathcal{A} = \{ \frac{j\pi}{4} | j=0,\dots,7 \}$.
            \item The server sends a single-qubit quantum state $\rho$.
        \end{itemize}
    }
    \Comp{
        \begin{enumerate}
            \item Set $\rho' = R_z(\theta) \rho (R_z(\theta))^\dagger$ and return $\rho'$ to the server.
        \end{enumerate}
    }
\end{resource}

Resource~\ref{res:bdqc} captures \emph{Blind Delegated Quantum Computing} without verifiability~\cite{dunjko2014composable}. In our version of the ideal functionality, we restrict the delegated computation to classical inputs. A malicious server (setting $c=1$) has the option to corrupt the state that the client receives at the end of the protocol.

\begin{resource}
    \caption{Blind Delegated Quantum Computing}
    \label{res:bdqc}
    \DontPrintSemicolon
    \SetKwInput{Inputs}{Inputs}
    \SetKwInput{Comp}{Computation by the resource}
    \Inputs{
        \begin{itemize}
            \item The client sends a classical bit string $x$.
            \item The server sends a flag $c \in \{0,1\}$. If $c=1$, the server also sends a quantum state $\psi$, and the description of a map $\mathcal{E}$.
        \end{itemize}
    }
    \Comp{
        \begin{enumerate}
            \item If $c=0$, the resource computes the correct output $y = \mathcal{U}(|x\rangle\langle x|)$ and sends it to the client.
            \item If $c=1$, the resource computes the corrupted output $y = \mathcal{E}(|x\rangle\langle x| \otimes \psi)$ and sends it to the client.
        \end{enumerate}
    }
\end{resource}

Note that universal computation is possible with Resource~\ref{res:bdqc} if $\mathcal{U}$ is a universal quantum map, and the description of the target quantum computation is encoded in the client's input $x$. For honest servers, the filter $\perp_{c=0}$ sets $c=0$ and blocks access to the other interface on the server's side.

Finally, we need to describe the ideal functionality of our target Resource~\ref{res:mcbqc}, \emph{Multi-Client Blind Quantum Computation}. As before, the jointly evaluated computations are restricted to classical inputs.

\begin{resource}
    \caption{Multi-Client Blind Quantum Computation}
    \label{res:mcbqc}
    \DontPrintSemicolon
    \SetKwInput{Inputs}{Inputs}
    \SetKwInput{Comp}{Computation by the resource}
    \Inputs{
        \begin{itemize}
            \item For $j=1,\dots,n$, client $j$ sends a classical bit string $x_j$. It also inputs $c_j \in \{0,1\}$ as a filtered interface.
            \item The server inputs a flag $c \in \{0,1\}$.
            \item All malicious parties, that is all clients with $c_j=1$, and the server if $c=1$, jointly send a quantum state $\psi$ and the description of a map $\mathcal{E}$.
        \end{itemize}
    }
    \Comp{
        \begin{enumerate}
            \item If $c_j=c=0$ for all $j=1,\dots,n$, the resource computes the correct output $y = \mathcal{U} \left( \bigotimes_{j=1}^n |x_j\rangle\langle x_j| \right)$ and sends it to the clients.
            \item Otherwise, the resource computes the corrupted output $y = \mathcal{E} \left( \bigotimes_{j=1}^n |x_j\rangle\langle x_j| \otimes \psi \right)$ and sends it to the clients.
        \end{enumerate}
    }
\end{resource}

\subsection{Security of the full protocol}

Universal Blind Quantum Computing~\cite{broadbent2009universal} is a protocol that allows a client to delegate a computation to a server with the guarantee that the server cannot learn anything about the computation, its inputs, or its outputs, except for the amount of resources that were used. In other words, only the computation graph and the order of qubits are leaked to the server. Protocol~\ref{prot:bqc} is a version of this protocol that does not require universal resource states, but rather works directly on the graph that is used to implement the target computation.

\begin{protocol}
    \caption{Blind Quantum Computation}
    \label{prot:bqc}
    \DontPrintSemicolon
    \SetKwInput{Public}{Public Information}
    \SetKwInput{Inputs}{Inputs}
    \SetKwInput{Protocol}{Protocol}
    \Public{
        \begin{itemize}
            \item A graph $G = (V, E, I, O)$ with input and output vertices $I$ and $O$, respectively.
            \item A partial order $\preceq$ on the set $V$ of vertices.
        \end{itemize}
    }
    \Inputs{
        \begin{itemize}
            \item The client has as input a classical bit string $x \in \{0,1\}^{|I|}$. It further inputs a set of angles $(\phi_v)_{v\in V}$, and a flow $f$ on $G$ compatible with $\preceq$.
            \item The server has no inputs.
        \end{itemize}
    }
    \Protocol{
        \begin{enumerate}
            \item For all $v\in V$, the client samples an angle $\theta(v) \gets_R \mathcal{A}$ uniformly at random, prepares the single-qubit state $\ket{+_{\theta(v)}}$, and sends it to the server.
            \item The server applies the entangling operation according to the graph $G$, i.e., for every edge $\{v,w\} \in E$, the server applies the two-qubit gate $\mathsf{CZ}_{vw}$.
            \item For all $v\in V \setminus O$, such that the partial order $\preceq$ is respected, the client and the server perform the following interactive steps:
                \begin{enumerate}
                    \item The client uses the previous (corrected) measurement outcomes to compute the corrected angle $\phi'(v)$ from $\phi(v)$ according to the flow $f$.
                    \item The clients samples a bit $r(v) \gets_R \{0,1\}$ uniformly at random, calculates $\delta(v) = \phi'(v) + \theta(v) + (r(v) + x(v)) \pi$, and sends $\delta(v)$ to the server.
                    \item The server measures qubit $v$ in the $\ket{\pm_{\delta(v)}}$-basis and returns the measurement outcome $m(v)$ to the client.
                    \item The client calculates the corrected measurement outcome as $m'(v) = m(v) \oplus r(v)$.
                \end{enumerate}
            \item The client outputs the bit string $(m'(v))_{v\in V \setminus O}$. The server further sends the qubits in $O$ to the client, who decrypts them (using $\theta(v)$, $r(v)$ and $m'(v)$ as keys) and keeps them as additional output.
        \end{enumerate}
    }
\end{protocol}

Previous work showed that Resource~\ref{res:rsr} can be used in combination with Protocol~\ref{prot:bqc} to securely construct Resource~\ref{res:bdqc}. Since we use this result as part of our security proof, we reiterate it here for convenience \cite[Theorem~4]{ma2022qenclave}.

\begin{theorem}\label{thm:rsr-bdqc}
    The BQC Protocol~\ref{prot:bqc} where the client uses Resource~\ref{res:rsr} to remotely prepare the required single-qubit states perfectly constructs Resource~\ref{res:bdqc} against a malicious server.
\end{theorem}

The security analysis of the multi-client protocol proposed in this work follows a modular paradigm. In this spirit, we first analyze the security of Protocol~\ref{prot:crsr}, the subprotocol which consists of the communication of a single qubit along the Qline.
Since both the photon source and the server are untrusted and potentially colluding in the protocol, they are treated as a single untrusted party in the following.

\begin{protocol}
    \caption{Collaborative Remote State Rotation}
    \label{prot:crsr}
    \DontPrintSemicolon
    \SetKwInput{Inputs}{Inputs}
    \SetKwInput{Protocol}{Protocol}
    \Inputs{
        \begin{itemize}
            \item The $n$ clients have no input.
            \item The orchestrator has as input an angle $\theta \in \mathcal{A}$.
            \item The server receives as input a single-qubit quantum state $\rho$.
        \end{itemize}
    }
    \Protocol{
        \begin{enumerate}
            \item For $j=1,\dots,n$, client $j$ samples uniformly at random $\theta_j \gets_R \mathcal{A}$, and sends $\theta_j$ to the orchestrator.
            \item The server sends $\rho$ to client 1.
            \item For $j=1,\dots,n$, client $j$ applies the operation $R_z(\theta_j)$ to the received quantum state and forwards it to the next client. After applying its own rotation, client $n$ forwards the final state to the server.
            \item The orchestrator computes $\theta' = \theta - \sum_{j=1}^n \theta_j \pmod{2\pi}$ and sends $\theta'$ to the server.
            \item The server applies the operation $R_z(\theta')$ to the single-qubit state that it received from client $n$, and keeps the resulting state as its output.
        \end{enumerate}
    }
\end{protocol}

We continue to prove the security of Protocol~\ref{prot:crsr}.

\begin{theorem}\label{thm:crsr}
    Protocol~\ref{prot:crsr} perfectly constructs Resource~\ref{res:rsr} between the orchestrator and the server from secure classical and quantum channels against malicious coalitions of at most the server and $n-1$ clients.
\end{theorem}

\begin{proof}[Proof of correctness.]
    If all participating parties are acting honestly, the single-qubit quantum state sent from client $n$ to the server takes the following form:
    \begin{align*}
        R_z(\bar\theta) \rho \left(R_z(\bar\theta)\right)^\dagger,
    \end{align*}
    where $\bar\theta = \sum_{j=1}^n \theta_j$. After the final correction which is applied to this state by the server, the output becomes
    \begin{align*}
        R_z(\theta') R_z(\bar\theta) \rho \left(R_z(\bar\theta)\right)^\dagger \left(R_z(\theta')\right)^\dagger = R_z(\theta) \rho \left(R_z(\theta)\right)^\dagger,
    \end{align*}
    which shows that the protocol is correct.
\end{proof}

\begin{proof}[Proof of soundness.]
    As security in the AC framework is simulation-based, we need to provide the construction of a simulator fit to translate real-world to ideal-world attacks. In the following, we assume the worst case of a colluding malicious coalition of the server and $n-1$ clients. Because the protocol and the ideal resource are sufficiently symmetric in the enumeration of the clients, we can assume without loss of generality that the first client behaves honestly. The construction of the simulator for this scenario is given as Simulator~\ref{sim:crsr}.

\begin{simulator}
    \caption{Malicious server and clients $2,\dots,n$}
    \label{sim:crsr}
    \DontPrintSemicolon
    \SetKwInput{Behavior}{Behavior of the simulator}
    \Behavior{
        \begin{enumerate}
            \item The simulator expects angles $\theta_j \in \mathcal{A}$ for $j=2,\dots,n$ from the malicious clients, and a single-qubit quantum state $\rho$ from the malicious server.
            \item The simulator forwards $\rho$ to the ideal functionality described by Resource~\ref{res:rsr}, and receives the state $\rho'$ from it.
            \item It samples uniformly at random the angle $\theta_1 \gets_R \mathcal{A}$.
            \item It applies the operation $R_z(\theta_1)$ to $\rho'$ and returns the resulting quantum state to the malicious server.
            \item Finally, the simulator computes $\theta' = - \sum_{j=1}^n \theta_j$ and sends $\theta'$ to the malicious server.
        \end{enumerate}
    }
\end{simulator}

    It remains to be shown that the views of the distinguisher in the real world where it has access to the inputs $\theta, \rho$ and to the views of all malicious parties, and in the ideal world where it has access to the inputs $\theta, \rho$ and to the interfaces to the simulator are perfectly equal. These two views can be summarized as follows:

\begin{center}
\begin{tabular}{lcc}
\hline
& \textbf{Real world} & \textbf{Ideal world} \\
\hline
\textit{Input angle} & $\theta$ & $\theta$ \\
\textit{Input state} & $\rho$ & $\rho$ \\ 
\textit{Client 1 output} & $R_z(\theta_1) \rho (R_z(\theta_1))^\dagger$ & $R_z(\theta + \theta_1) \rho (R_z(\theta + \theta_1))^\dagger$ \\
\textit{Correction} & $\theta - \sum_{j=1}^n \theta_j$ & $-\sum_{j=1}^n \theta_j$ \\
\hline
\end{tabular}
\end{center}

    Since $\theta_1$ is chosen uniformly at random by the simulator in the ideal world, we can substitute it by $\theta_1 - \theta$ without changing the view of the distinguisher. This yields:

\begin{center}
\begin{tabular}{cc}
\hline
\textbf{Real world} & \textbf{Ideal world} \\
\hline
$\theta$ & $\theta$ \\
$\rho$ & $\rho$ \\ 
$R_z(\theta_1) \rho (R_z(\theta_1))^\dagger$ & $R_z(\theta_1) \rho (R_z(\theta_1))^\dagger$ \\
$\theta - \sum_{j=1}^n \theta_j$ & $\theta - \sum_{j=1}^n \theta_j$ \\
\hline
\end{tabular}
\end{center}

    Clearly, these two distributions are identical, which proves that the views of the distinguisher in the two worlds are perfectly indistinguishable.
\end{proof}

It now remains to piece together all building blocks to obtain the security of the full protocol.

\begin{theorem}
    The Multi-Client Blind Quantum Computation Protocol~1 of the main text, where the orchestrator is replaced by a classical SMPC resource between all other parties, perfectly constructs Resource~\ref{res:mcbqc} against malicious coalitions of at most the server and $n-1$ clients.
\end{theorem}

\begin{proof}
    After having established the security of all building blocks, this proof is straightforward and works by repeated application of the general composition principle in Theorem~\ref{thm:ac-composition}, analogously to the proof of~\cite[Theorem~5]{kapourniotis2023asymmetric}.

    Retracing the steps of the construction of the protocol in question, we begin by replacing the classical SMPC resource by a trusted classical party, the orchestrator. This step does not incur any security loss.

    Next, Theorem~\ref{thm:crsr} allows us to replace every execution of Protocol~\ref{prot:crsr} by one call to Resource~\ref{res:rsr}, with zero security loss.

    Finally, Theorem~\ref{thm:rsr-bdqc} establishes that the remaining protocol between orchestrator and server is indeed a secure construction of the Blind Delegated Quantum Computing Resource~\ref{res:bdqc}. Keeping in mind that the orchestrator additionally is in charge of collecting the clients' inputs and distributing the outcome of the computation, this protocol is indeed a perfect realization of Resource~\ref{res:mcbqc}.
\end{proof}

\begin{remark}[Removing redundant correction steps]
    When considering the realization of a resource by itself, the final correction step in Protocol~\ref{prot:crsr} is indeed necessary, as this is the only way for the simulator to transmit the correct quantum state to the server.
    However, when using the protocol in the context of UBQC, this correction step can be combined with the corrections that are anyway present in the BQC protocol, and do hence not need to be performed as a separate round of communication.
    This observation is similar to the one made by~\cite{kapourniotis2023asymmetric} in the context of Collaborative RSP and QSMPC.
\end{remark}

\begin{remark}[Varying the location of the entangling step]
    The security of the protocol is independent of whether the qubits are entangled before or after their communication along the Qline. Therefore, both setups in which the photon source creates cluster states and the server performs only adaptive measurements, as well as setups in which the photon source emits unentangled qubits, and entanglement and measurements are performed by the server are possible. Even combinations of both, in which part of the entanglement is created before, and part after the Qline, are conceivable.
\end{remark}

\section{Time scheme of the experimental protocol}\label{sec:time_scheme}

At the beginning of each round of the protocol, the clients Alice and Bob set their liquid crystals (LCs) according to the phase shifts they want to apply to their qubits (the $\theta_i^j$ random parameters) and communicate all their secret parameters to the system dedicated to the TTP. With this information, the TTP computes the measurement angle for the first photon according to the formula $\delta_1$, and it pre-computes the two possible values for $\delta_2$, given that the outcome of the first measurement is still unknown, namely $\delta_2^{\pm} = \theta_2 + x_2 \pi + r_2\pi \pm \phi_2 $. These angles correspond to two possible voltages, either $V_1$ or $V_2$. This stage is necessary to keep the time delay between the two photons reasonable, and so the losses. Indeed, while the first round of classical communication would take time intervals of the order of the milliseconds (since one would need to wait for all the settings to be performed), the communication of the measurement outcome alone takes, instead, some tens of nanoseconds only, since it happens through analog signals.  $\approx$ ms time-delays would need hundreds of km long-fibers, while $\approx$ ns time-delays only needs some meters long fibers.
The first measurement angle $\delta_1$ is communicated to $S_2$. Starting from time $t_1$, the quantum part of the experiment begins.
The photon pairs are generated and sent through Alice's and Bob's LCs, and $S_2$ measures the first qubit. The outcome of this measurement, $m_1$, is communicated to the TTP. Starting from $t_2$, the TTP can communicate the measurement angle $\delta_2$ to $S_2$ through a voltage that  activates suitably the Pockels cell, and so the second photon is measured. In the end, the outcome of this measurement is sent to the TTP which communicates the final result of the measurement to the clients, namely $m^{true}_2 = m_2 \oplus r_{2} $.

\begin{figure}[h!]
    \centering
    \includegraphics[width=\textwidth]{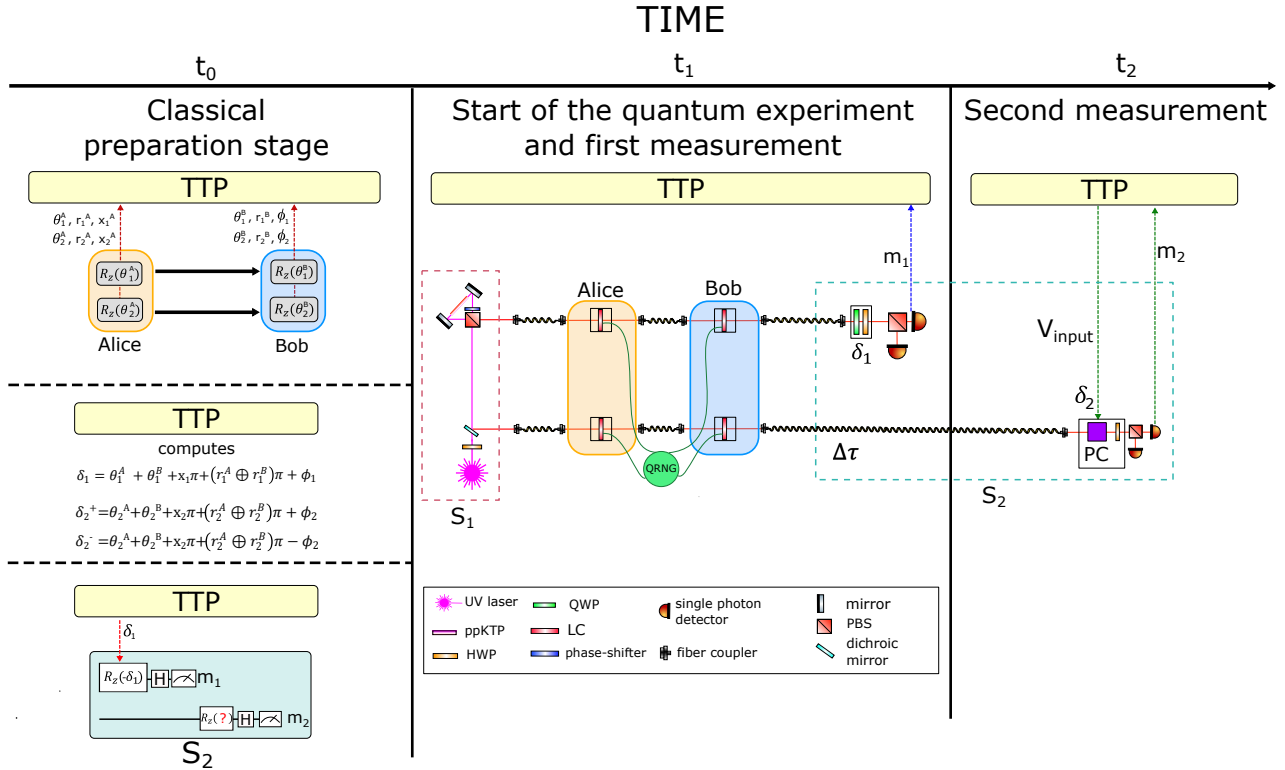}
    \caption{\textbf{Time scheme of the experiment.} Here, we depict the time scheme of the experiment. It is divided into three main stages: ($t_0$) the classical preparation stage, where the clients set up their liquid crystals and communicate their secret parameters to the TTP, ($t_1$) the start of the quantum part of the protocol and the first measurement, ($t_2$) third and last round where the second measurement happens. The second photon is delayed through a $\approx 65$~m long single-mode fiber.}
    \label{fig:time_scheme}
\end{figure}

\section{State preparation and measurement}\label{sec:state_preparation}

We encode the computational basis in the polarization of photons, namely $\ket{0} = \ket{H}$ and $\ket{1} = \ket{V}$.
We generate entangled pairs of photons in the state $\frac{1}{\sqrt{2}}\left(\ket{HH}+\ket{VV} \right)$ and then suitably rotate the polarization of the second photon to obtain the state $\frac{1}{2}\left( \ket{HH} + \ket{HV} + \ket{VH} - \ket{VV}\right)$.

Afterward, the clients rotate their photons in the desired state, namely, by randomly choosing the parameters $\theta_1^A,~\theta_2^A,~\theta_1^B,~\theta_2^B$, from the set $\{0,~\pi/4,...,~7\pi/4 \}$, and by manipulating its polarization through a sequence of liquid crystals, whose action can be described by the following operator:

\begin{equation}
    R_{z}(\theta) = 
    \begin{pmatrix}
        1 & 0 \\
        0 & e^{i\theta}
    \end{pmatrix}
\label{eq:Rz_theta}
\end{equation}

Projections in the basis $\{ \ket{+_{\delta_1}}, \ket{-_{\delta_1}} \}$ are performed by use of a quarter-wave plate rotated at angle $-\pi/4$ with respect to its optic axis, followed by a half-wave plate (HWP) rotated at angle $\pi/8+\delta_1/4$ with respect to its optic axis, and a polarizing beam splitter (PBS).

\begin{equation}
    O(\delta_1) = 
    U^{\dagger}_{QWP}(-\pi/4)~U^{\dagger}_{HWP}(\pi/8+\delta_1/4)~O_{PBS}~U_{HWP}(\pi/8+\delta_1/4)~U_{QWP}(-\pi/4)
\label{eq:O_delta1}
\end{equation}

Projections in the basis $\{ \ket{+_{\delta_2}}, \ket{-_{\delta_2}} \}$ are performed by use of a Pockels cell which is activated with a suitable voltage, corresponding to a given phase $\phi$, followed by a HWP rotated at angle $\pi/8$ with respect to its optic axis, and a PBS, as described by Eq.~\eqref{eq:O_delta2}.

In Table~\ref{tab:phase-shifts}, we show the phase shifts necessary to implement the desired measurement $O(\delta_2)$. We optimize the number of different voltages needed, by applying an outcome flip ($f$) whenever the phase shift required is larger or equal to $\pi$. The resulting operator is the following:
\begin{equation}
    O(\delta_2) = 
    X^f~U^{\dagger}_{PC}(\phi)~U^{\dagger}_{HWP}(\pi/8)~O_{PBS}~U_{HWP}(\pi/8)~U_{PC}(\phi)~X^f
\label{eq:O_delta2}
\end{equation}

\begin{table}[h!]
\centering
\begin{tabular}{cccc}
$\boldsymbol{\delta_2}$ & \textbf{Ideal phase-shift} & \textbf{PC phase-shift} & \textbf{f} \\
0                   & 0                          & 0                       & 0                    \\
$\pi/4$             & $7\pi/4$                   & $3\pi/4$                & 1                   \\
$\pi/2$             & $3\pi/2$                   & $\pi/2$                 & 1                   \\
$3\pi/4$            & $5\pi/4$                   & $\pi/4$                 & 1                   \\
$\pi$               & $\pi$                      & 0                       & 1                   \\
$5\pi/4$            & $3\pi/4$                   & $3\pi/4$                & 0                    \\
$3\pi/2$            & $\pi/2$                    & $\pi/2$                 & 0                    \\
$7\pi/4$            & $\pi/4$                    & $\pi/4$                 & 0                   
\end{tabular}
\caption{}
\label{tab:phase-shifts}
\end{table}

\section{System for measurement adaptivity}\label{sec:adaptivity}

In order to experimentally implement the feed-forward system, i.e. measuring the first qubit and consequently choosing the measurement angle $\delta_2$ according to the outcome $m_1^{true}$, it is necessary to introduce a temporal delay between the two photons' paths of $\approx 320$~ns or $\approx 65$~m in fiber. This time delay corresponds to the time needed to compute $\delta_2$ and to feed the Pockels cell with a suitable voltage accordingly.
Supplementary Table~\ref{tab:phase-shifts} shows that the eight possible values for $\delta_2$ can be implemented by four independent rotations, performed by a Pockels cell, in addition to a final bit-flip ($f=0,1$) performed in post-processing by the server.

We designed a logic circuit that takes as input nine bits and produces a five-bit output, in formulas

\begin{equation}
    g(A,B,r_1,m_1^+,m_1^-)=VT
\end{equation}

Specifically, the three-bit inputs $A=a_2a_1a_0 \text{~and~} B=b_2b_1b_0\in\{000,001,\dots,111\}$ encode respectively the clients' chosen parameters $\theta_2+x_2\pi +r_2 \pi \text{~and~} \phi_2\in\{0,\tfrac{\pi}{4},\dots,\tfrac{7\pi}{4}\}$. The single-bit $r_1 = r_1^A\oplus r_1^B$ corrects the value of $m_1$ into $m_1^{true}$. The two-bit input $\{m_1^+,~m_1^- \}\in\{10,01\}$ represents the only interesting physical output of the two detectors in the server's first measurement station encoding respectively $m_1=0,1$. We define the function $g$ such that, if $m_1^{true} = 0(1)$, then $\delta_2 = A +(-) B$.
Thus, according to the theory, the inputs $r_1,~m_1^+,~m_1^-$ let the device choose if either to perform the sum or difference, both modulo eight, of the inputs $A,~B$, that corresponds to evaluating the angle $\delta_2$. 
Instead, the three-bit output $V=v_2v_1v_0$ encodes the evaluated $\delta_2$ angle, while the output bits $m_1^{+true},~m_1^{-true}\in\{10,01\}$ encode respectively $m_1^{true}=0,1$.
A scheme of our device is shown in Fig.~\ref{fig:fast data elaboration}.
Our encoding values for $A,~B,~f,~V$ are shown in the Supplementary Table~\ref{tab:FDE-encodings}, in addition to the association between the output $fV$ and the required Pockels cell voltage $\text{V}_i$ for $i=0,\tfrac{\pi}{4},\tfrac{\pi}{2},\tfrac{3\pi}{4}$.
\begin{table}[h!]
\centering
\begin{tabular}{llrl}
$a_2a_1a_0\leftrightarrow A[\text{rad}]\quad$ & $b_2b_1b_0\leftrightarrow B[\text{rad}]\quad$ & $\delta_2\leftrightarrow fv_2v_1v_0$ & $\leftrightarrow$\textbf{PC voltage} \\
$\quad000\qquad0$                             & $\quad000\qquad0$                             & $0\qquad0000\quad$                   & $\qquad\text{V}_0$ \\
$\quad001\qquad\pi/4$                         & $\quad001\qquad\pi/4$                         & $\pi/4\qquad1100\quad$               & $\qquad\text{V}_{3\pi/4}$ \\
$\quad010\qquad\pi/2$                         & $\quad010\qquad\pi/2$                         & $\pi/2\qquad1010\quad$               & $\qquad\text{V}_{\pi/2}$ \\
$\quad011\qquad3\pi/4$                        & $\quad011\qquad3\pi/4$                        & $3\pi/4\qquad1001\quad$              & $\qquad\text{V}_{\pi/4}$ \\
$\quad100\qquad\pi$                           & $\quad100\qquad\pi$                           & $\pi\qquad1000\quad$                 & $\qquad\text{V}_0$ \\
$\quad101\qquad5\pi/4$                        & $\quad101\qquad5\pi/4$                        & $5\pi/4\qquad0100\quad$              & $\qquad\text{V}_{3\pi/4}$ \\
$\quad110\qquad3\pi/2$                        & $\quad110\qquad3\pi/2$                        & $3\pi/2\qquad0010\quad$              & $\qquad\text{V}_{\pi/2}$ \\
$\quad111\qquad7\pi/4$                        & $\quad111\qquad7\pi/4$                        & $7\pi/4\qquad0001\quad$              & $\qquad\text{V}_{\pi/4}$
\end{tabular}
\caption{}
\label{tab:FDE-encodings}
\end{table}

\begin{figure}
    \centering
    \includegraphics[width=0.47\textwidth]{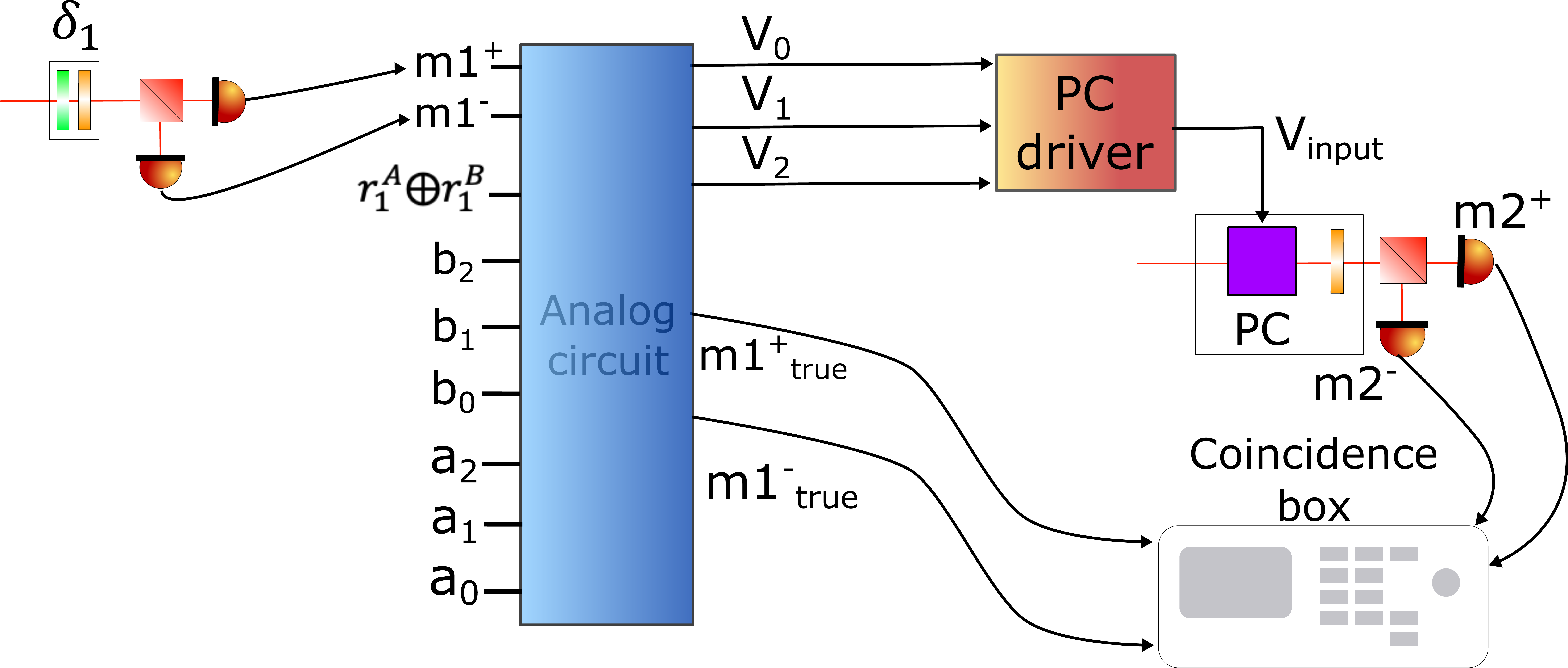}
    \caption{\textbf{Fast data elaboration system.} A suitably programmed analog circuit gets as input: (i) three bits corresponding to the following angle $A = a_2a_1a_0 = (\theta_2+ x_2 + r_2\pi)\%(2\pi) \in \{0,...,7\pi/4 \}$, (ii) three bits corresponding to $B = b_2b_1b_0 = \phi_2$, (iii) one bit equal to $r_1 = (r_1^A \oplus r_1^B)$, (iv) two signals corresponding to the outcome 0 and 1 of the first measurement, respectively $m_1^+$ and $m_1^-$. The function first corrects the outcome of the first measurement according to the formula: $m_1^{\pm true} = m1^{\pm} \oplus r_1$. Then, it computes the second measurement basis $\delta_2$ and transforms it into a low-voltage pulse to be amplified and then applied to the Pockels cell. In detail, the outcome of the circuit is made up of: (i) three bits, $V_0,~V_1,~V_2$, corresponding to three different voltage amplification factors performed by the PC driver, (ii) a copy of the two corrected signals $m1^{\pm}_{true}$. The HV voltage pulse is sent to the PC and, after the second measurement is performed, its outcome is sent to a coincidence box as well as the first one. The second outcome is corrected in post-processing, by applying the formula $m2^{\pm}_{true} =  m2^{\pm} \oplus f \oplus r_2$, where the values of $f$ can be retrieved in Table~\ref{tab:phase-shifts}.}
    \label{fig:fast data elaboration}
\end{figure}

\section{Blindness of the first qubit}\label{sec:blindness_first}
Here, we report the details of the demonstration of the blindness of the first qubit. To show it, we fixed the measurement angle $\delta_2 = 3\pi/2$ and averaged the resulting density matrices of the first qubit over all possible clients' rotation angles $\theta_2^A$ and $\theta_2^B$, namely $64$ possible combinations. Here, as an instance, we report the quantum state tomographies for eight combinations, where Alice keeps her rotation angle fixed at $\theta_2^A = 0$, while Bob varies it. The quantum state tomographies shown also confirm the correctness of the protocol in the case of the computation of a quantum function, taking the first qubit as the output of the computation. Indeed, each of these density matrices can be seen as the case where the computation angle amounts to $\phi_2 = \delta_2 -\theta_2^A-\theta_2^B $, where $\delta_2 = 3 \pi / 2$ and  $\theta_2^A = 0$.

\begin{figure}[H]
    \centering
    \includegraphics[scale=0.38]{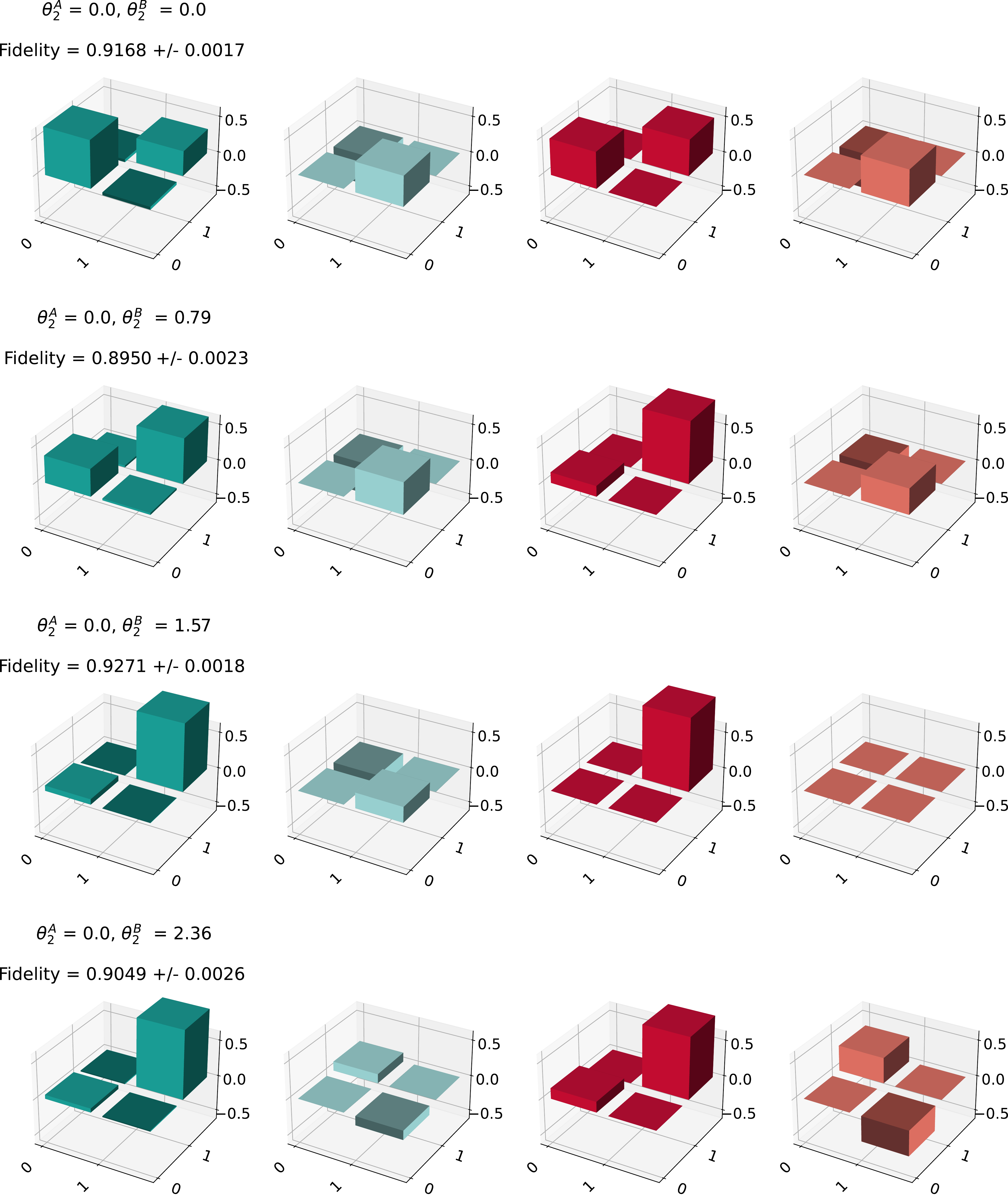}
    
\end{figure}

\begin{figure}[H]
    \centering
    \includegraphics[scale=0.38]{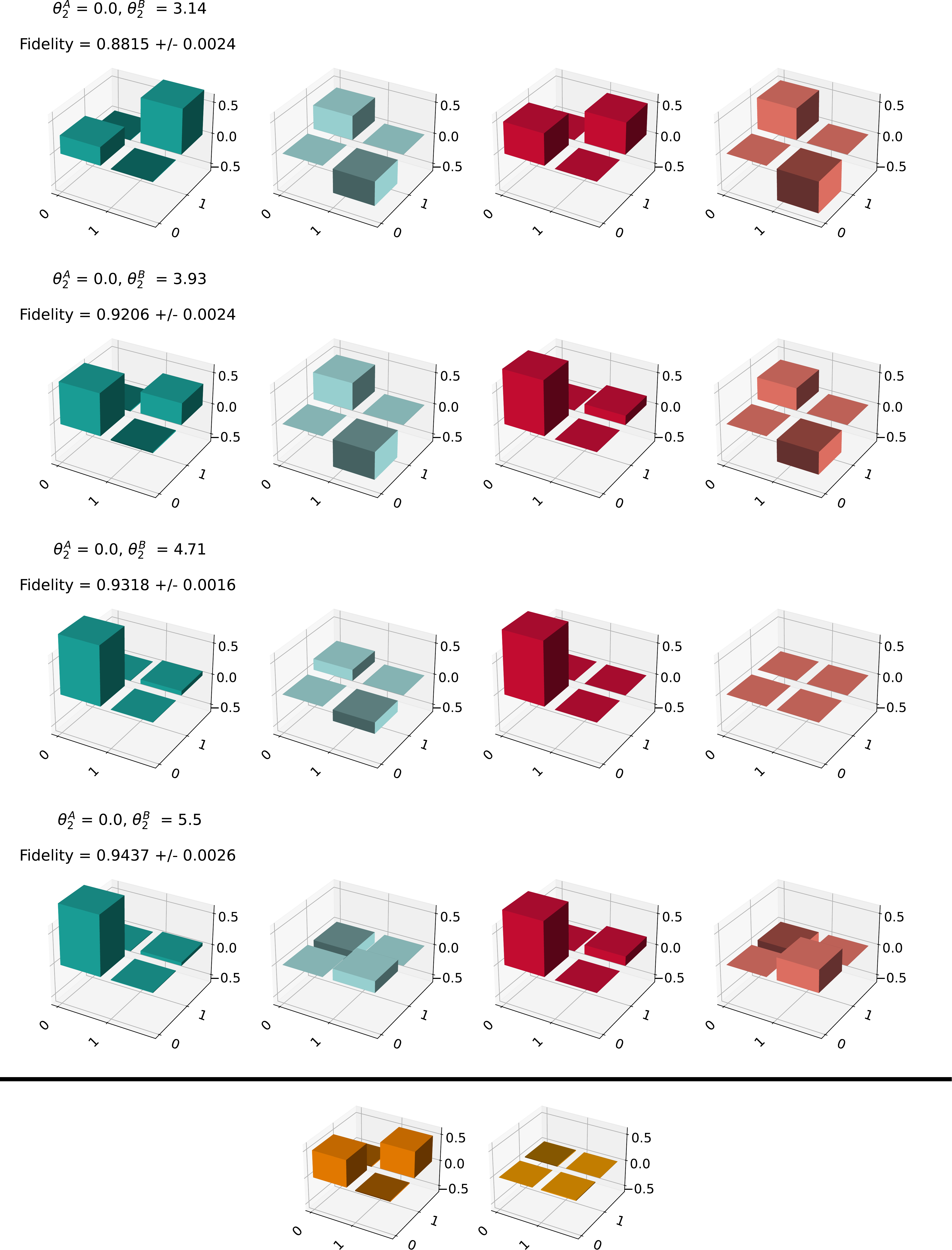}
    \caption{\textbf{Experimental and theoretical quantum state tomographies to demonstrate blindness of the first qubit.}  We show in blue and light blue, respectively, the real and imaginary parts of the experimental density matrix, and in red and light red the real and imaginary parts of the ideal density matrix. We also report the corresponding fidelities and the resulting completely mixed state. The fidelity with the completely mixed state amounts to $F_1 = 0.99949 \pm 0.00002$ and the Von Neumann entropy amounts to $S_1 = 0.99852 +/- 0.00006$.}
    \label{fig:tomo_blindness}
\end{figure}

\section{Blindness of the second qubit}\label{sec:blindness_second}

In this section, we report the details of the demonstration of the blindness of the second qubit. To show it, we fixed the measurement angle $\delta_1 = \pi$ and averaged the resulting density matrices for the second qubit over all possible clients' rotation angles of the first qubit, namely $64$ possible combinations. Here, as an instance, we report the quantum state tomographies for eight combinations and their comparison with the theoretical expectations, where Alice keeps her rotation angle fixed at $\theta_1^A = \pi$, while Bob varies it. The quantum state tomographies shown also confirm the correctness of the protocol in the case of the computation of a quantum function. Indeed, each of these density matrices can be seen as the case where the first computation angle amounts to $\phi_1 = \delta_1 -\theta_1^A-\theta_1^B $, where $\delta_1 = \pi$ and  $\theta_1^A = \pi$.

\begin{figure}[H]
    \centering
    \includegraphics[scale=0.39]{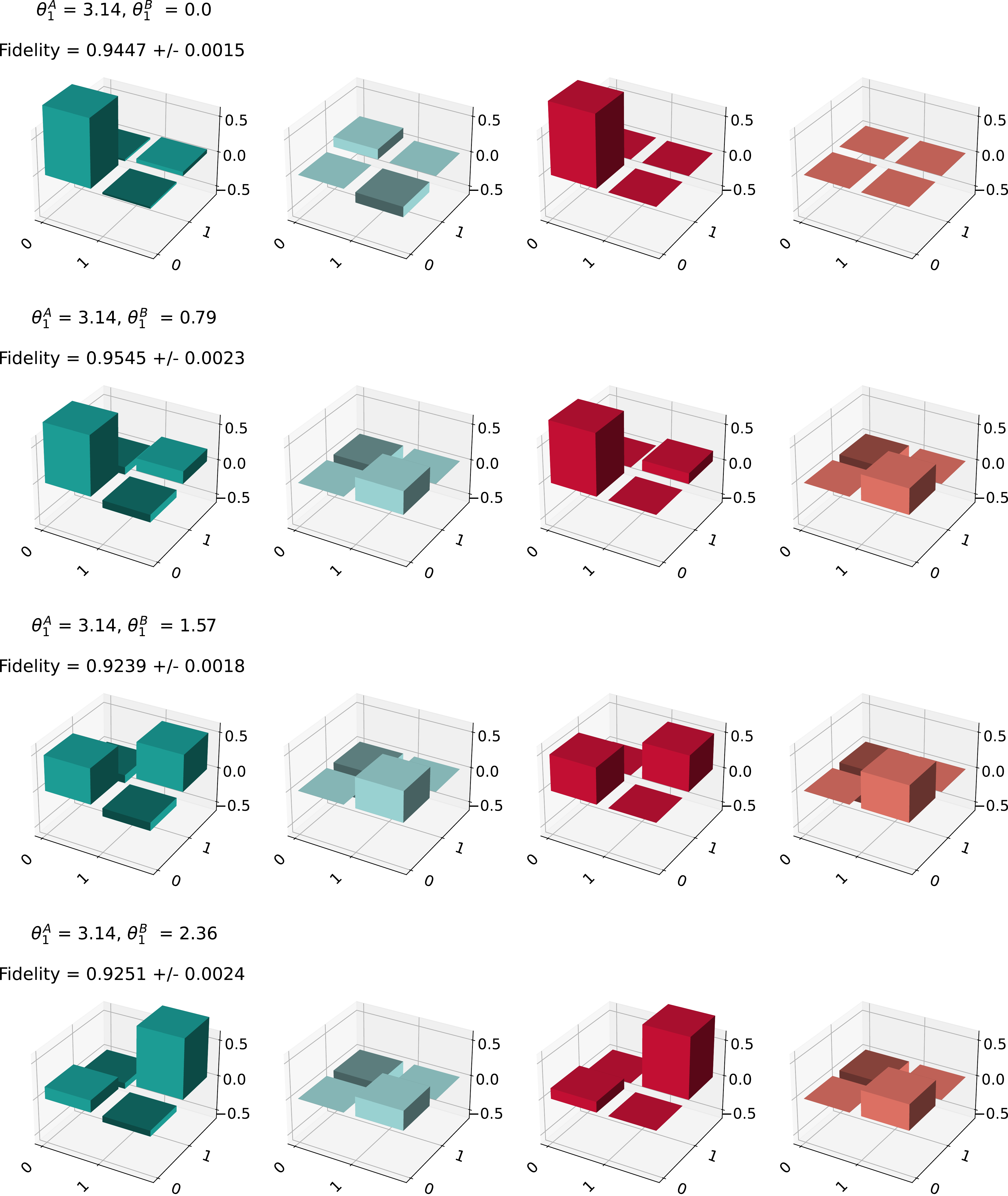}
    
\end{figure}

\begin{figure}[H]
    \centering
    \includegraphics[scale=0.39]{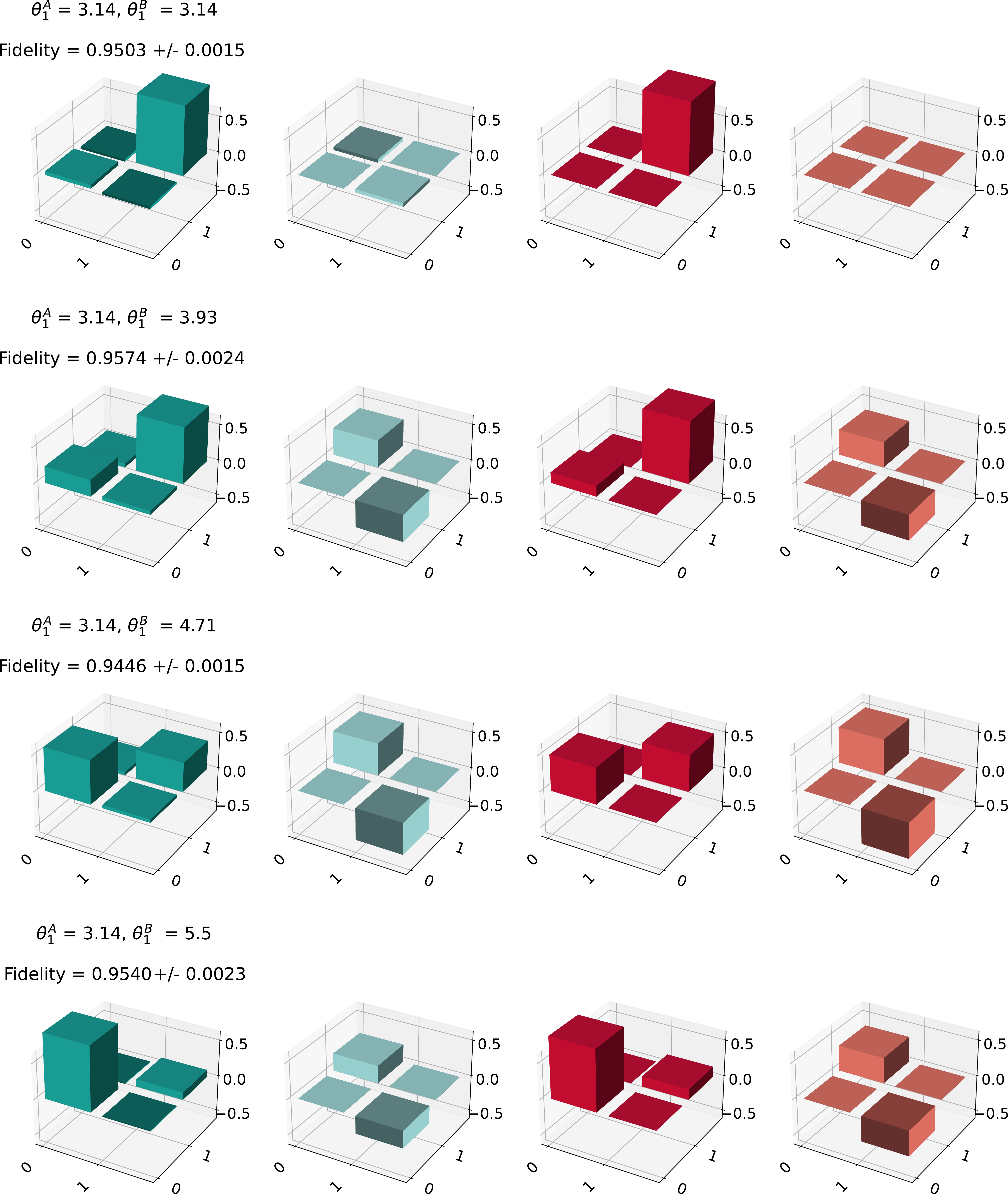}
    \caption{\textbf{Experimental and theoretical quantum state tomographies to demonstrate blindness of the second qubit.} We show in blue and light blue, respectively, the real and imaginary parts of the experimental density matrix, and in red and light red the real and imaginary parts of the ideal density matrix. We also report the corresponding fidelities.}
    \label{fig:tomo_blindness_secondo}
\end{figure}

\section{Blindness of the full initial state}\label{sec:blindness_both}

In this section, as an example, we report the quantum state tomographies for the first $16$ combinations of the clients' parameters $\theta_1^A,\theta_2^B$ and the comparison with the theoretical expectations. We demonstrated the blindness of the full initial state by averaging over all density matrices resulting from varying $\theta_1^A,~\theta_2^B$ in the set $\mathcal{A}$.

\begin{figure}[H]
    \centering
    \includegraphics[scale=0.42]{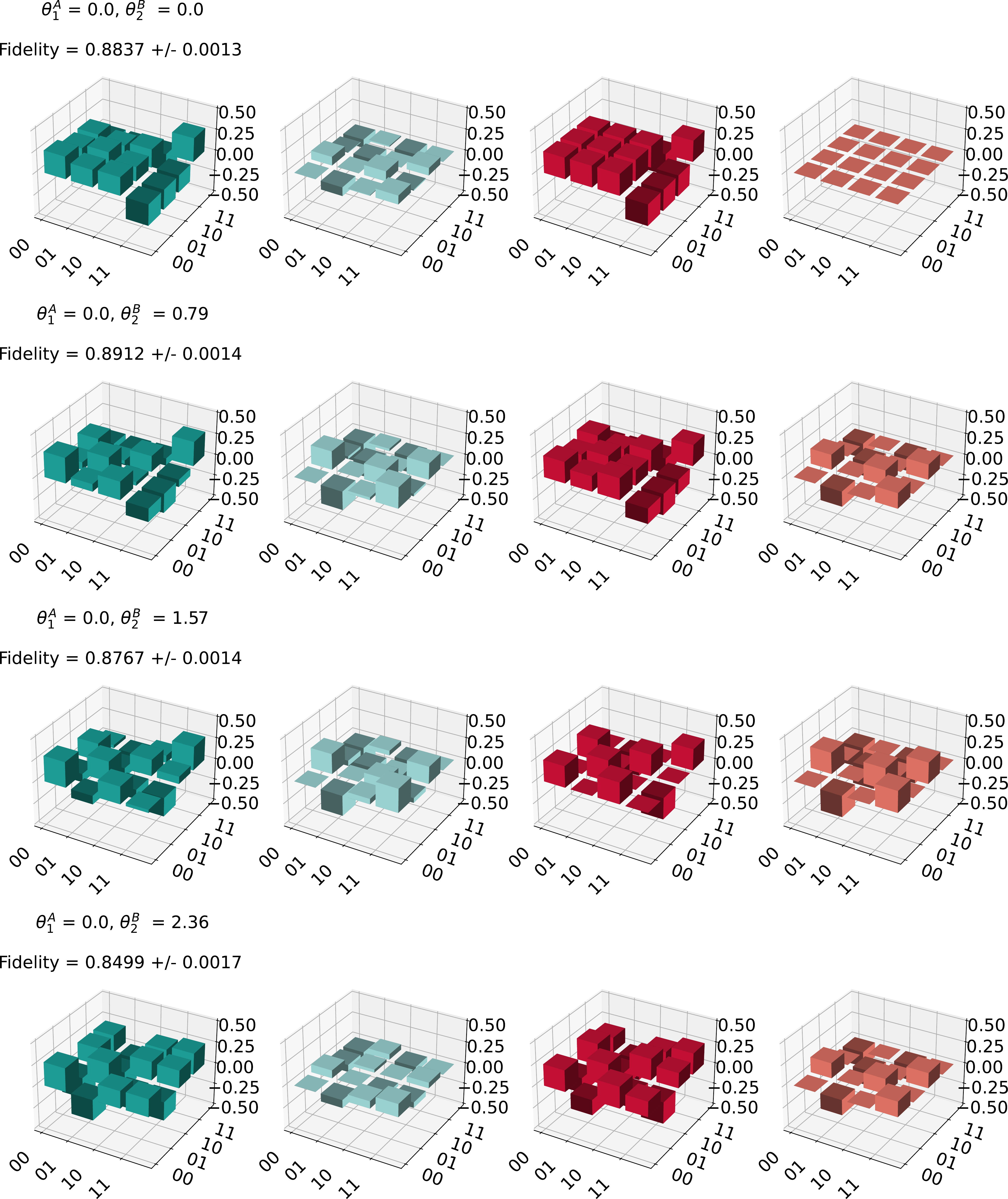}
\end{figure}

\begin{figure}[H]
    \centering
    \includegraphics[scale=0.42]{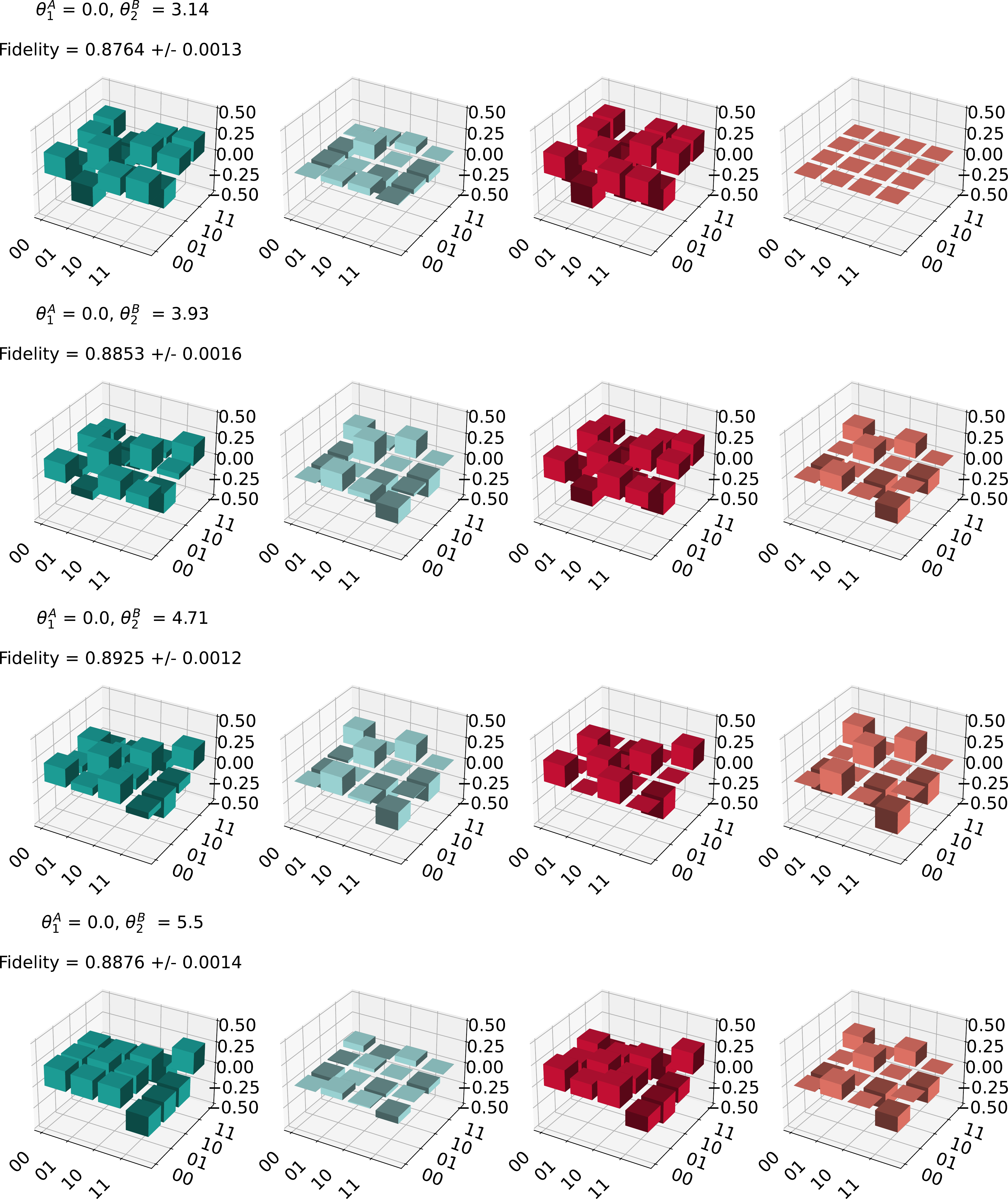}
\end{figure}

\begin{figure}[H]
    \centering
    \includegraphics[scale=0.42]{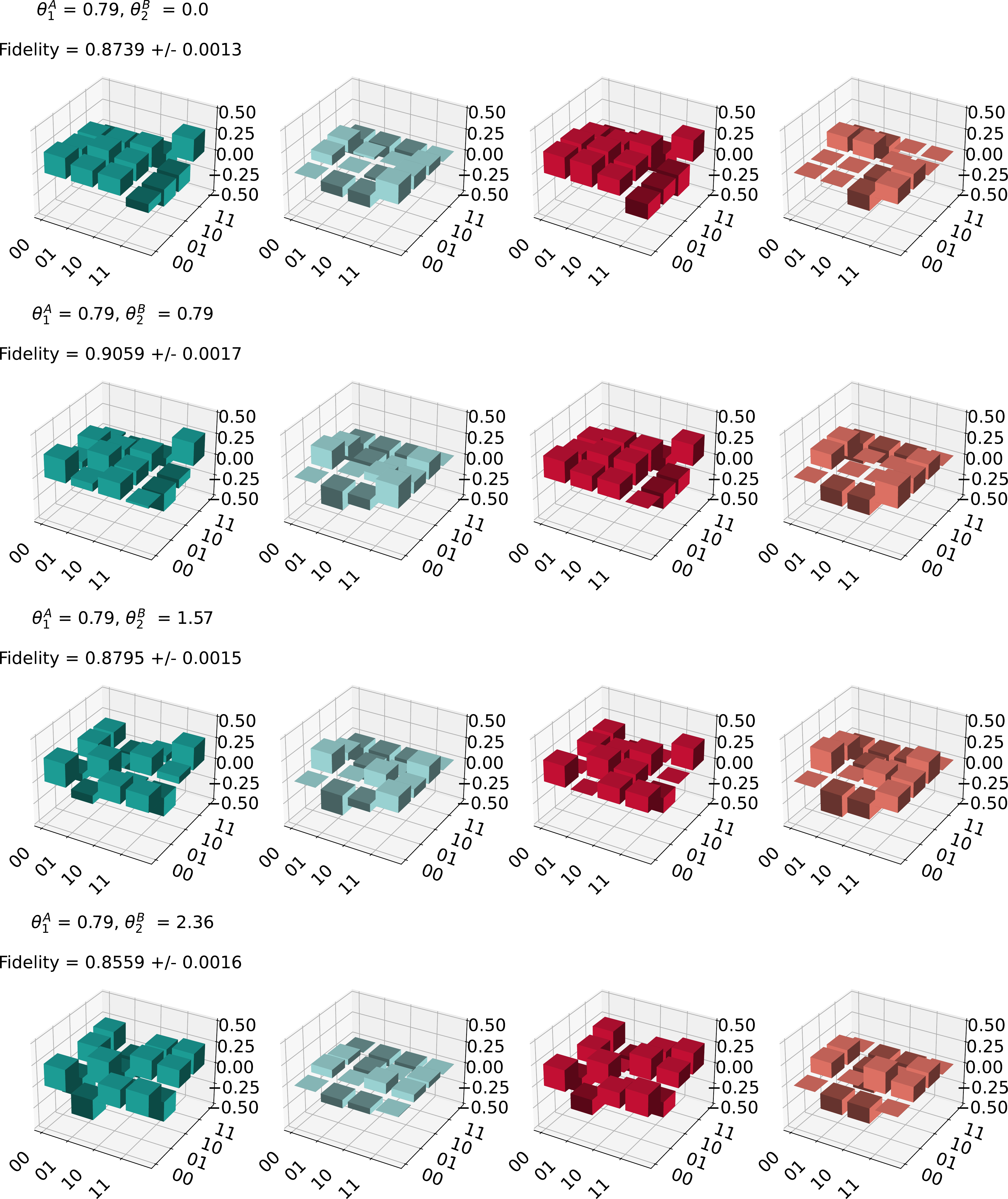}
\end{figure}

\begin{figure}[H]
    \centering
    \includegraphics[scale=0.42]{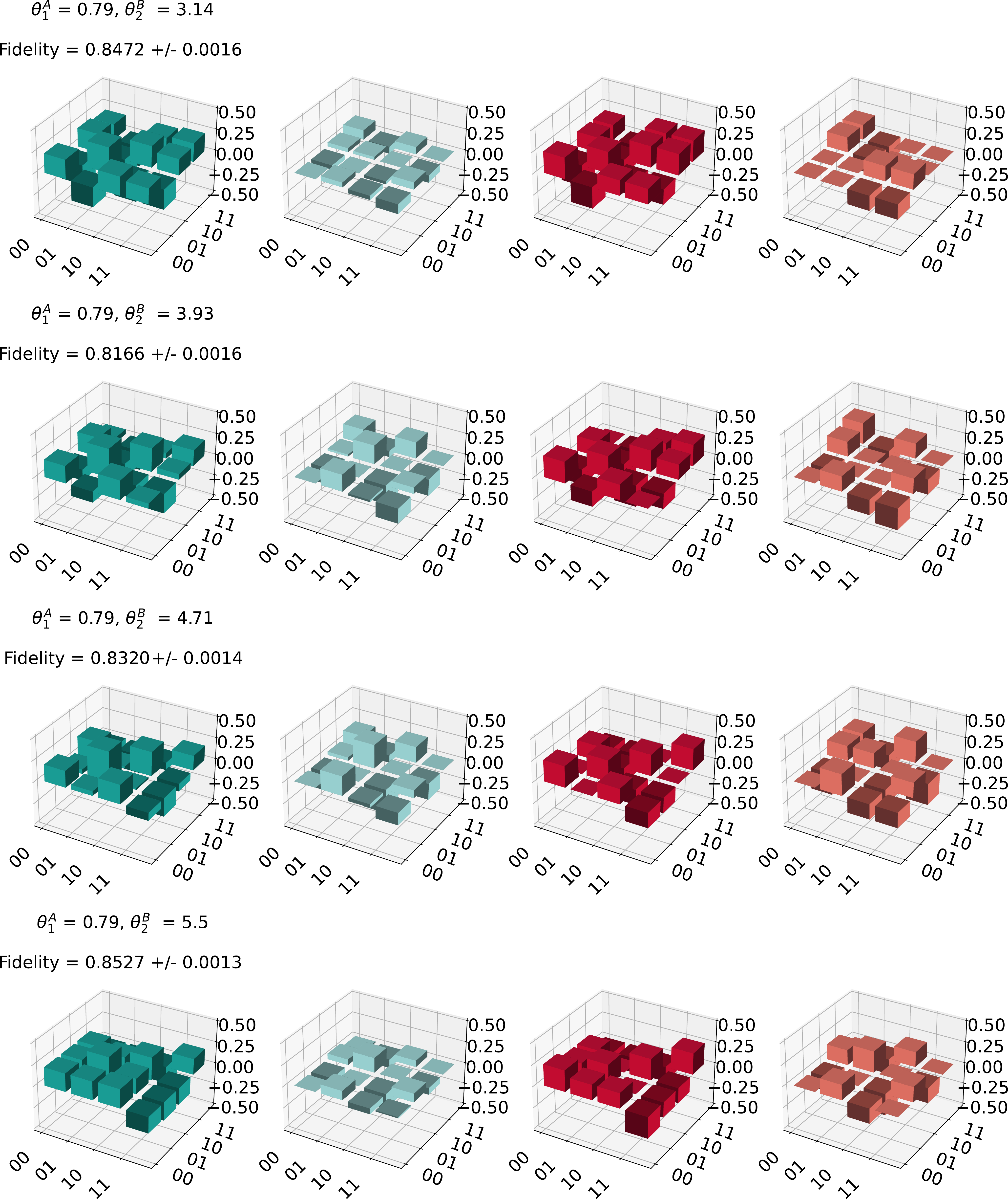}
    \caption{\textbf{Experimental and ideal quantum state tomographies of $16$ different initial two-qubit states.} We show in blue and light blue, respectively, the real and imaginary parts of the experimental density matrix, and in red and light red the real and imaginary parts of the ideal density matrix. We also report the corresponding fidelities.}
\end{figure}

\section{Noise model}\label{sec:noise_model}

Ideally, the state generated by the source is an entangled state in the form:

\begin{equation}
    \ket{\psi^-} = \frac{1}{2} \left( \ket{HH}   + \ket{HV} + \ket{VH} - \ket{VV} \right)
\end{equation}

We modeled the experimental state by introducing white and colored noise, according to the following expression:

\begin{equation}
    \rho_{noisy} = v \ket{\psi^-}\bra{\psi^-} + (1-v) \left[ \frac{\lambda}{2}~( \ket{\psi^+}\bra{\psi^+} + \ket{\psi^-}\bra{\psi^-} ) + \frac{1-\lambda}{4} \mathbb{I}_4 \right ]
\end{equation}
where $v$ is the visibility of the state and $l$ is the fraction of coloured noise and $\ket{\psi^+} = \frac{1}{2} \left( \ket{HH}   + \ket{HV} - \ket{VH} + \ket{VV} \right)$.

The parameters $v$ and $\lambda$ compatible with our data are $v \approx 0.76$ and $\lambda \approx 0.63$, on average for all states generated by varying the liquid crystals.

We also consider the imperfect phase shift inserted by the Pockels cell, which is on average $\approx -\pi/20$ with respect to the expected phase shift.

\section{Correctness of the protocol}\label{sec:correctness}

\begin{figure}[httb]
    \includegraphics[width=0.6\textwidth]{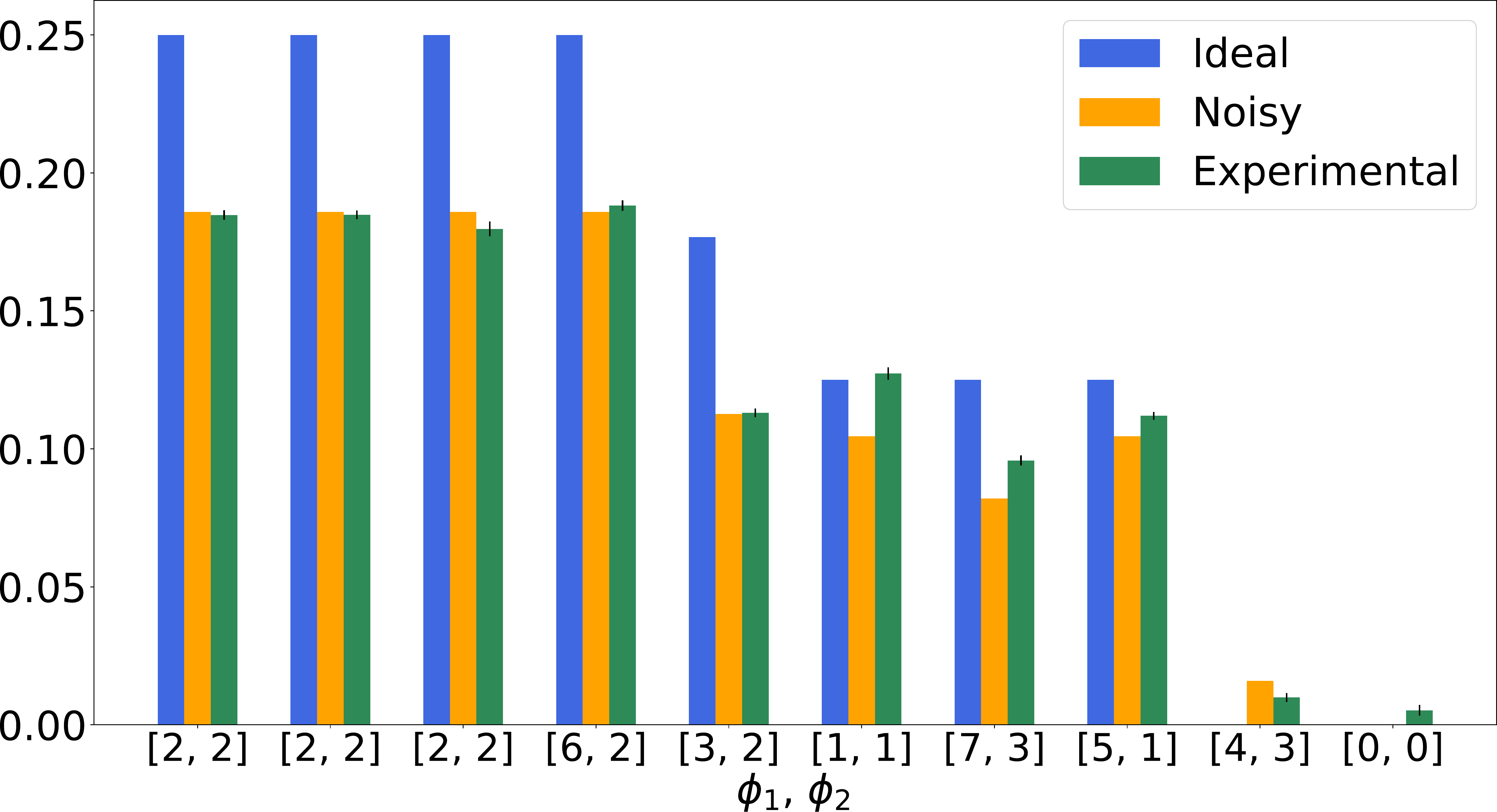}
    \caption{\textbf{Expected average distance from the uniform distribution in the ideal, noisy-modeled, and experimental case}. In this bar plot, we show the expected average distance from the uniform distribution in the ideal case, in a model including experimental noise, and in the experimental case, for the algorithms considered in the main text. Uncertainties on the experimental data were estimated by assuming Poissonian statistics. Black bars correspond to one standard deviation.}
    \label{fig:ideal_distance_from_uniform}
\end{figure}

Here, we report further details about the correctness of the protocol. In Fig.~\ref{fig:ideal_distance_from_uniform}, we show the comparison between the average distance from the uniform distribution in the ideal, noisy-modeled case and experimental case for the algorithms considered in the Results section of the main text. We define the average distance between two distributions $\mathbf{p}$ and $\mathbf{q}$ as follows, where $p_i,~q_i$ are the distribution elements and $N$ is the total number of elements ($N=4$ for our distributions):
\begin{equation}
    d(\mathbf{p}, \mathbf{q}) = \frac{\sum_i |p_i - q_i|}{N}
    \label{eq:ave_distance}
\end{equation}
We chose as a figure of merit the distance from the uniform distribution since this is the distribution one would have if the computation resource was classical. Therefore, the plot in Fig.~\ref{fig:ideal_distance_from_uniform} provides additional validation of the nonclassicality of the computational resource.

\begin{figure}[httb]
    \includegraphics[width=0.8\textwidth]{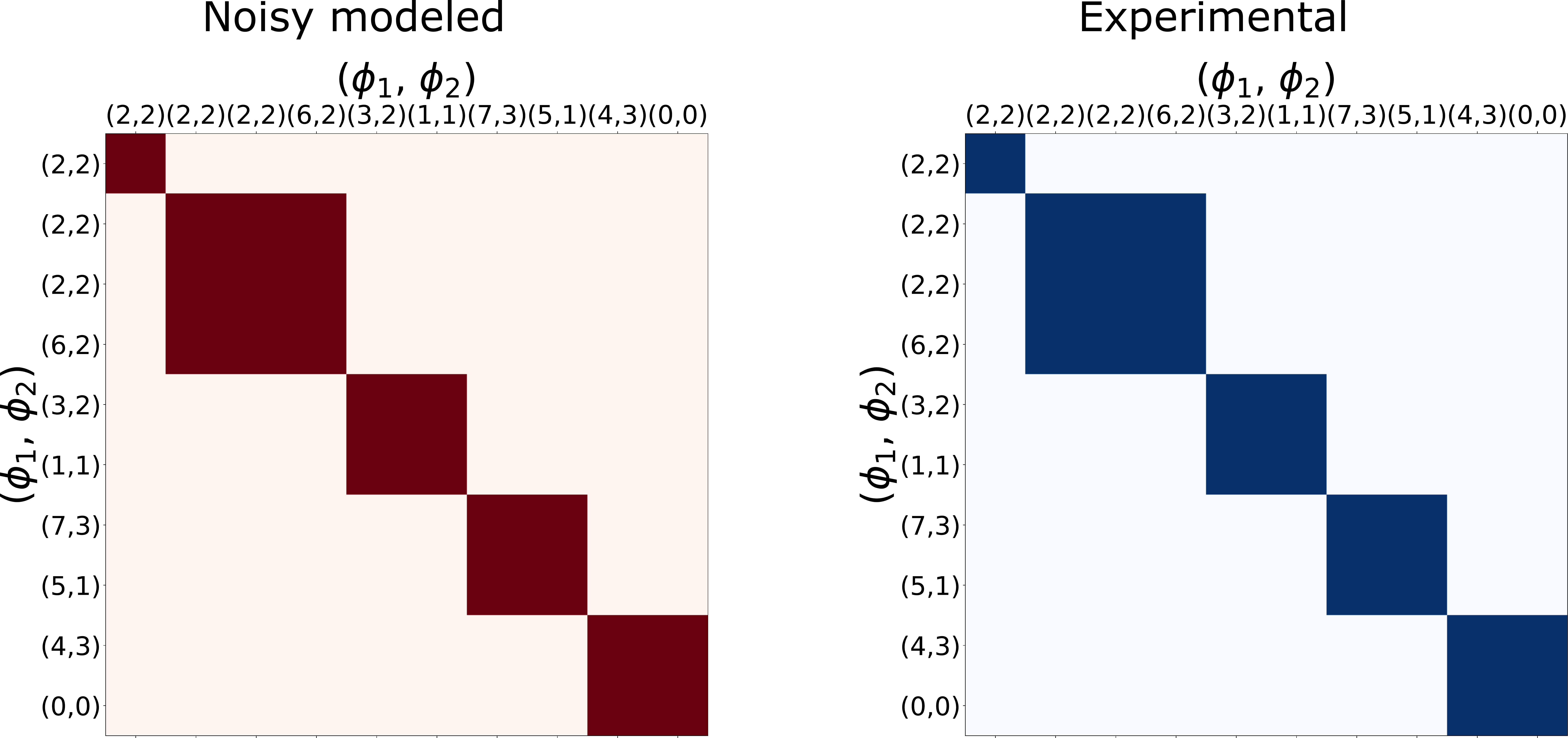}
    \caption{\textbf{Average distance between noisy-modeled and experimental frequencies}. In these color matrices, we analyze the expected and measured overlaps between the algorithms considered in this work. In particular, in the red plot, we show the expected distances in the noisy model, while, in the blue one, we show the distance between experimental and noisy modeled frequencies.
    We highlight in dark color the cases in which the average distance between the distributions on the $x$-axis and $y$-axis is lower than or equal to $0.05$. The light-colored parts, instead, correspond to distances greater than $0.05$. The pairs $\left( \phi_1, \phi_2 \right)$ are indicated as integer multiples of $\frac{\pi}{4}$.}
    \label{fig:matrix_distance_between_probs}
\end{figure}

In Fig.~\ref{fig:matrix_distance_between_probs}, we show that, by comparing each experimental distribution with each noisy-modeled one for all algorithms, the best overlap is with their noisy-modeled counterparts. This means that each algorithm can be distinguished from the others within a distance $\approx 0.05$.

